\documentclass[structabstract]{aa} 
\usepackage{graphicx}
\usepackage{txfonts}
\usepackage{natbib}
\bibpunct{(}{)}{;}{a}{}{,} 

\begin{document}
\title{Gas and dust in the TW Hydrae Association as seen by the Herschel Space Observatory\thanks{{\it Herschel} is an ESA space observatory with
    science instruments provided by European-led Principal
    Investigator consortia and with important participation from
    NASA.}}
   \author{P. Riviere-Marichalar\inst{1,5}, C. Pinte\inst{2}, D. Barrado\inst{1,4}, W. F. Thi\inst{2}, C. Eiroa\inst{5},  I. Kamp\inst{6}, B. Montesinos\inst{1}, J. Donaldson\inst{7}, J.C. Augereau\inst{2}, N. Hu\'elamo\inst{1}, A. Roberge\inst{8}, D. Ardila\inst{9}, G. Sandell\inst{10},  J. P. Williams\inst{11},W. R. F. Dent\inst{12}, F. Menard\inst{2,3}, J. Lillo-Box\inst{1},G. Duch\^{e}ne\inst{2,13}
}

   \institute{Centro de Astrobiolog\'{\i}a -- Depto. Astrof\'isica (CSIC--INTA), ESAC Campus, P.O. Box 78, 
	     28691 Villanueva de la Ca\~nada, Spain\\
             \email{riviere@cab.inta-csic.es, barrado@cab.inta-csic.es}
             \and UJF-Grenoble 1 / CNRS-INSU, Institut de Plan\'{e}tologie et d'Astrophysique (IPAG) UMR 5274, Grenoble, F-38041, France 
             \and Laboratorio Franco-Chileno de Astronomia (UMI 3386: CNRS -- U de Chile / PUC / U Conception),  Santiago, Chile 
             \and Calar Alto Observatory, Centro Astron\'{o}mico Hispano-Alem\'{a}n C/Jes\'{u}s Durb\'{a}n Rem\'{o}n, 2-2, 04004 Almer\'{i}a, Spain 
             \and Dep. de F\'isica Te\'orica, Fac. de Ciencias, UAM Campus Cantoblanco, 28049 Madrid, Spain 
               \and Kapteyn Astronomical Institute, P.O. Box 800, 9700 AV Groningen, The Netherlands 
             \and Department of Astronomy, University of Maryland, College Park, MD 230742, USA 
             \and Exoplanets and Stellar Astrophysics Lab, NASA Goddard Space Flight Center, Code 667, Greenbelt, MD, 20771, USA 
             \and NASA Herschel Science Center, California Institute of Technology, MC 100-22, Pasadena, CA 91125, USA 
             \and SOFIA-USRA, NASA Ames Research Center 
             \and Institute for Astronomy, University of Hawaii, Honolulu, USA 
             \and ALMA, Avda Apoquindo 3846, Piso 19, Edificio Alsacia, Las Condes, Santiago, Chile 
             \and Astronomy Department, University of California, Berkeley CA 94720-3411 USA 
}
   \authorrunning{Riviere et al.}
   \date{}

 \abstract
{Circumstellar discs are the places where planets form, therefore knowledge of their evolution is crucial for our understanding of planet formation. The \textit{Herschel Space Observatory} is providing valuable data for studying disc systems, thanks to its sensitivity and wavelength coverage. This paper is one of several devoted to analysing and modelling \textit{Herschel}-PACS observations of various young stellar associations from the GASPS Open Time Key Programme.}
{The aim of this paper is to elucidate the gas and dust properties of circumstellar discs in the $\rm \sim$ 10~Myr TW~Hya Association (TWA) using new far-infrared (IR) imaging and spectroscopy from \textit{Herschel}-PACS.}
{We obtained far-IR photometric data at 70, 100, and 160 $\rm \mu m$ of 14 TWA members; spectroscopic observations centred on the [OI] line at 63.18 $\rm \mu m$ were also obtained for 9 of the 14. The new photometry for each star was incorporated into its full spectral energy distribution (SED).
}
{We detected excess IR emission that is characteristic of circumstellar discs from five TWA members, and computed upper limits for another nine. 
Two TWA members (TWA~01 and TWA~04B) also show [OI] emission at 63.18~$\mu {\rm m}$. Discs in the TWA association display a variety of properties, with a wide range of dust masses and inner radii, based on modified blackbody modelling. Both transitional and debris discs are found in the sample. Models for sources with a detected IR excess give dust masses in the range from $\rm \sim 0.15 M_{\oplus}$ to $\rm \sim 63 M_{\oplus}$.}  
{}
   \maketitle
\section{Introduction} 
We can distinguish several phases in the evolution of circumstellar discs. In the protoplanetary stage, the star actively accretes material from the disc, which is gas-rich and extends up to hundreds of AU from the star. This protoplanetary disc is the birthplace of planets. In the transition phase, the disc forms an inner gap that is largely dust-free \citep{Strom1989,Najita2007}. The radius of the gap is typically in the range 1--20 AU, and may be the result of interactions between the disc and a planetary body \citep{Lin1993}. At this stage, the star may still be accreting, but accretion rates are 10 to 100 times lower than in the protoplanetary phase. Debris discs are older systems that are gas-poor and no longer accreting. The presence of dust in such evolved systems is explained as the result of destructive collisions between planetesimals formed earlier in the protoplanetary phase. In this paradigm, dusty circumstellar discs surrounding stars are signposts for the existence of planetesimals and possibly of planets themselves. 

The properties of circumstellar discs have been extensively studied over the past 30 years or so \citep[see recent reviews by e.g.][and references therein]{Krivov2010,Williams2011}. The Multiband Imaging Photometer for Spitzer (MIPS) was widely used to study the properties of dusty discs in evolved systems \citep[e.g.][]{Bryden2006,Rebull2008}. The \textit{Spitzer} Infrared Spectrograph (IRS) was also used to characterise dust grains in circumstellar environments \citep[e.g.][]{Chen2006,Morales2009,Oliveira2011}. By fitting models to the spectral energy distribution (SED) of a disc system, the amount of dust, the geometry of the disc, the dust composition, and the grain size distribution can be estimated. Even though the gas dominates the mass in a protoplanetary disc, it is more difficult to study than dust, because $\rm H_{2}$ lacks a permanent electric dipole moment, which causes the IR pure rotational transitions to be very weak. A further difficulty is that molecules like CO are often optically thick and frozen out in the cold mid-plane of the disc, making them poor tracers of the overall gas content.

The \textit{Herschel Space Observatory} \citep{Pilbratt2010} provides new opportunities for studies of circumstellar discs. The \textit{Photodetector Array Camera \& Spectrometer} (PACS) \citep{Poglitsch2010} has unprecedented sensitivity and angular resolution in the far-IR, enabling the detection of gas and dust in faint, dusty debris discs in the wavelength range where the cold Kuiper-Belt like dust emission peaks (70-200 $\mu \rm{m}$). The smaller beam size of PACS compared with \textit{Spitzer}-MIPS results in a lower rate of confusion with background sources. PACS spectroscopy covers the strong cooling lines of [OI] at 63 $\mu \rm{m}$ and [CII] at 158 $\mu \rm{m}$, which provides a powerful probe of the gas in circumstellar discs.

The discovery of several nearby ($\lesssim$ 100 pc) young ($\lesssim$ 20 Myr) sparse stellar associations or moving groups \citep{Torres2003,Torres2006} provides another valuable avenue for studying the evolution of gas and dust in circumstellar discs, because the ages of the stars in these moving groups are relatively well known. In the present paper, we target the TW Hydrae Association \citep[TWA, ][]{deLaReza1989}. TWA is one of the closest (56 pc) and youngest moving groups \citep[$\rm 8-20~Myr$, see ][]{Kastner1997,Stauffer1995,Soderblom1998,Hoff1998,Weintraub2000,Makarov2001,Makarov2005,Barrado2006,deLaReza2006}. 
The range in ages is due to different age determinations from different authors, although we assume the TWA members are coeval. It is a well known place for studying star and planet formation, and the properties of the TWA circumstellar discs have been extensively studied in other works \citep[e.g.][]{Low2005}. 

Currently there are about 30 bona fide members of the TWA association \citep{Schneider2012}. Fourteen of them were observed as a part of the \textit{Herschel} Open Key Time Programme \textit{``Gas in Protoplanetary Systems''} \citep[GASPS; ][accepted]{Dent2013}, which aims to study the presence and distribution of gas and dust in circumstellar disc systems around young low- and intermediate-mass stars. GASPS has observed $\sim$ 250 stars within 400 allocated hours, covering both photometry and spectroscopy. The sample includes several associations at different ages, from Taurus at 1--3 Myr to Tucana-Horologium at $\rm \sim$ 30 Myr. The targets range from HAeBe stars to T~Tauri stars, with spectral types from A0 to M6. 

In this study, we discuss our new \textit{Herschel}-PACS photometric and spectroscopic observations of 14 TWA members. Images were obtained at 70 and/or 100, and 160 $\mu \rm{m}$ for all 14. Spectral line observations of [OI] at 63~$\mu \rm{m}$ and $\rm DCO^{+}$ at 189~$\mu \rm{m}$ were obtained for nine of them. 
We studied the properties of the dust and the geometry of the discs by comparing the observed photometric flux densities with those predicted by simple, modified blackbody models.
This paper is one of several studies devoted to the analysis of GASPS data in a systematic framework, with the goal of comparing the \textit{Herschel} photometric and spectroscopic results in different associations and moving groups, and studying the evolution of gas and dust as a function of age.

\section{The sample}
\begin{figure}[t]
\begin{center}
     \includegraphics[scale=0.5]{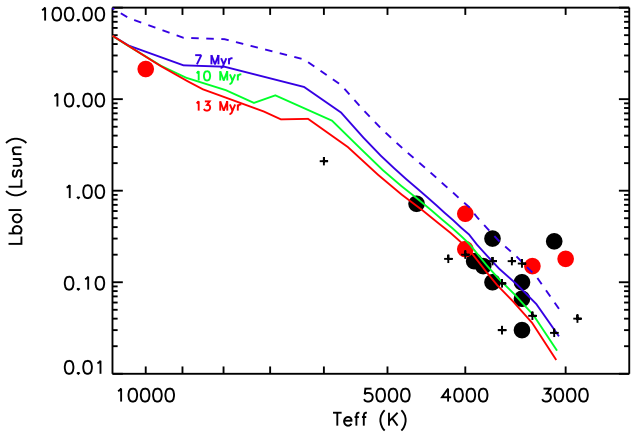}
   \caption{Hertzsprung-Russell diagram for TWA members in this programme. The red/green/blue solid lines represent the 13/10/7~Myr solar metallicity isochrones from \citet{Baraffe1998}. The blue dashed line shows the 7~Myr Baraffe isochrone with 2 times higher luminosity to account for unresolved binary systems. Red dots are objects detected with PACS, while black dots are undetected objects. Plus symbols correspond to TWA members not observed in GASPS.}
   \label{TWA_HRD}
\end{center}
\end{figure}

The sample of 14 TWA members studied in this paper, together with their spectral types (Sp Type), stellar luminosities, and effective temperatures, is shown in Table \ref{tableStar}. The spectral types range from A0 to M3. Star distances are taken from \cite{Zuckerman2004}. 
In Fig.~\ref{TWA_HRD}, we plot the stars on a Hertzsprung-Russell diagram. 
The stellar luminosities and effective temperatures used are discussed in section \ref{StParam}.  

Among the sample, TWA 01 \citep{Salyk2007}, TWA 03A \citep{Cieza2008}, and TWA 04B \citep{Furlan2007} are typically classified as transitional discs, while TWA 07 \citep{Matthews2007} and TWA 11A \citep{Telesco2000} are believed to be debris discs. It was proposed that \object{TWA~13A} harbours a circumstellar disc by \cite{Low2005}, but this was later rejected by \cite{Plavchan2009}. The eight other  systems are not known to harbour any circumstellar disc.

Archival data was collected for each TWA member in our target list, including Johnson, Tycho, 2-Microns All-Sky Survey (2MASS), InfraRed Array Camera (IRAC), Wide-Field Infrared Explorer (WISE), AKARI, Multiband Imaging Photometer for Spitzer (MIPS), Sub-Millimeter Array (SMA) and Submillimetre Common-User Bolometer Array (SCUBA) photometry. \textit{Spitzer}-MIPS values are taken from \cite{Low2005}. TWA23 and TWA25 MIPS flux densities are not included in \cite{Low2005}. For these two stars, we computed our own photometry for the 24 $\mu \rm{m}$ MIPS band, using an aperture of 13 arcsec, and a sky annulus between 20 arcsec and 32 arcsec, and applying the appropriate aperture corrections (as described in the \textit{Spitzer} Data Analysis Cookbook\footnote{http://ssc.spitzer.caltech.edu/dataanalysistools/cookbook/}). Computed flux densities are $\rm 8.6 \pm 0.5~mJy$ for TWA 23 and $\rm 10.5 \pm 0.5~mJy$.

\subsection{Stellar parameters}\label{StParam}
To model the stars' SEDs, we need knowledge of their stellar properties (specifically, their temperature and luminosity), to choose the correct photosphere models. For each star in the sample, we selected the photometric data that do not show excess above the stellar photosphere, i.e.\ those data points in agreement with pure photospheric emission. Typically, this includes Johnson, Stromgren, and 2MASS data. 
Then we compared this photospheric emission with a grid of theoretical stellar photosphere models using the Virtual Observatory SED Analyzer \citep[VOSA,][]{Bayo2008}, which provides the best-fitting model based on a $\chi^{2}$ minimisation. We used the grid of Phoenix models from \cite{Hauschildt1999}. The values for $\rm T_{eff}$ and $\rm L_{*}$ from each best-fitting model are summarised in Table~\ref{tableStar}. 

\begin{table}
\caption{Stellar parameters}             
\label{tableStar}      
\centering          
\begin{tabular}{l l l c l l l }     
\hline\hline       
Name & Other name  & Sp~type & Ref & d & T$\rm _{eff}$ &  L$\rm _*$  \\ 
\hline
	& & & & (pc) & (K)  & (L$_{sun}$)  \\ 
\hline                    
\object{TWA~01} & TW Hya & M2.5 & 1 &  56 & 3400  &	0.23 \\
\object{TWA~02} & CD-29 8887 & M2 & 2 & 52 & 3700  & 0.30  \\
\object{TWA~03A} & Hen 3-600 & M3 & 2 & 42 & 3000 & 0.18 \\
\object{TWA~04B} & HD 98800B & M5 & 3 & 47 & 4000 & 0.56 \\
\object{TWA~05A} & CD-33 7795 & M3 & 3 & 50 & 3100 & 0.28  \\
\object{TWA~07}  & CE Ant & M1 & 4 & 38 & 3300 & 0.15 \\
\object{TWA~10}  & V1252 Cen & M2.5 & 4  & 57 & 3400 & 0.10  \\
\object{TWA~11A} & HR 4796A  & A0 & 5  & 67 & 10000 & 21.3  \\
\object{TWA~12}  & V1217 Cen & M2 & 6  & 32 & 3400 & 0.03  \\
\object{TWA~13A} & V547 Hya A & M2 & 6 & 38 & 3700 & 0.10  \\ 
\object{TWA~16} & -- & M1.5 & 7  & 66 & 3900 & 0.17  \\
\object{TWA~21} & HD 298936  & K3 & 8 & 69  & 4600 & 0.72 \\
\object{TWA~23} & -- & M1 & 9  & 37 & 3400 & 0.07  \\
\object{TWA~25} & V1249 Cen  & M0 & 9 & 44 & 3800 & 0.15  \\
\hline                  
\end{tabular} 
\tablefoot{All distances are from \cite{Zuckerman2004}. $\rm T_{eff}$ and $\rm L_{*}$ for each target were computed using VOSA \citep{Bayo2008}. References for spectral type (Sp~Type) are: (1) \cite{Vacca2011}; (2) \cite{deLaReza1989}; (3) \cite{Gregorio1992}; (4) \cite{Webb1999}; (5) \cite{Houk1982}; (6) \cite{Sterzik1999}; (7) \cite{Zuckerman2001}; (8) \cite{Zuckerman2004}; (9): \cite{Weinberger2004}}
\end{table}

\subsection{Accretion in TWA}\label{Acc}
The relation between $\rm H_{\alpha}$ emission, a tracer of disc material accreting onto the central star, and [OI] emission can help us understand the gas properties of the TWA members. We took observed $\rm  H_{\alpha}$ emission equivalent widths from \cite{Barrado2006} and applied the criterion given in \cite{Barrado2003} to classify TWA members as accretors or non-accretors.  The accretion criterion accounts for the effect of chromospheric activity, which also gives rise to $\rm  H_{\alpha}$ emission.

\begin{figure}[!h]
\begin{center}
     \includegraphics[scale=0.53]{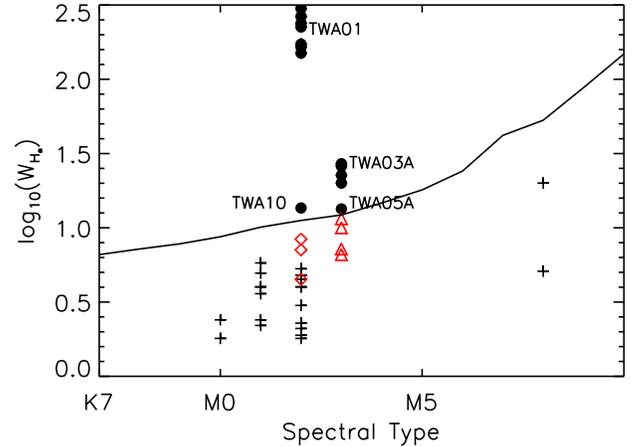}       
   \caption{Accretion in TWA members. Plus symbols depict objects with ${\rm H}_{\alpha}$ in agreement with pure chromospheric emission, while black dots are objects with ${\rm H}_{\alpha}$ in agreement with ongoing accretion. TWA 05A and TWA 10 show variable emission with some epochs over the chromospheric level and some epochs below the chromospheric level. Red diamonds and red triangles represent ${\rm H}_{\alpha}$ equivalent widths below the chromospheric level for TWA 10 and TWA 05A, respectively. The solid line shows the saturation criterion by \cite{Barrado2003}.} 
   \label{AccretionPlot}
\end{center}
\end{figure}

In Fig. \ref{AccretionPlot}, we have plotted the $\rm H_{\alpha}$ emission equivalent widths (EW) versus the spectral types of the stars in the sample, along with the line that separates accretors from non-accretors according to the criterion in \cite{Barrado2003}. \object{TWA~01} and \object{TWA~03A} show variable $\rm H_{\alpha}$ levels in agreement with an accreting disc. On the other hand, for \object{TWA~05A} and \object{TWA~10}, only one epoch of data shows an $\rm H_{\alpha}$ equivalent width larger than the expected chromospheric level.  We note that these stars show MIPS 24~$\rm \mu m$ emission consistent with purely photospheric emission, indicating a lack of warm inner material that would be expected for an accreting disc.  We therefore conclude that the two high $\rm H_{\alpha}$ observations were due to extremely strong flaring events rather than episodic accretion.

\section{Observations and data reduction}
The TWA sample was observed with PACS as part of the GASPS Open Time Key programme. Fourteen bona fide TWA members were observed in photometric mode and a subset of nine TWA members were observed in spectral line mode.
\subsection{Photometric data reduction}
The PACS photometer simultaneously observes in either the 70 $\rm \mu m$ or 100 $\mu \rm{m}$ bands together with the 160 $\mu \rm{m}$ band, so we typically have at least two images to combine in the 160 $\mu \rm{m}$ band. PACS scan map observations at 70 $\rm \mu m$ and 160 $\mu \rm{m}$ were obtained for 12 objects. Seven of these objects were also observed at 100 $\rm \mu m$ and 160 $\mu \rm{m}$. Two additional objects were observed only at 100 $\rm \mu m$ and 160 $\mu \rm{m}$. Observation IDs are listed in Table \ref{ObsLog}. The exposure times range from 133 s to 1122 s, based on the expected flux density from the star. Each scan map was made with medium speed ($\rm 20 \arcsec s^{-1}$), with scan legs of 3 $\arcmin$ and cross scan length of 4$\rm \arcsec$ to 5 $\rm \arcsec$. 

The reduction of the photometric data was carried out using the Herschel Interactive Processing Environment (HIPE) 8, with the latest version available for the calibration files. For bright IR sources (i. e., those with flux densities greater than 100 mJy) we used a version of the pipeline tuned for bright objects, while for faint objects and non-detected objects we used a different version optimised for noise-dominated maps, as is the case for TWA 07, among others. The difference between both versions of the pipeline comes from the way in which high-pass filtering is implemented. Before we apply high-pass filtering, we need to mask the sources that we want to study in order to preserve their absolute flux density. For bright sources, we can detect and mask the sources in individual scans before performing the high-pass filtering. For faint sources that cannot be detected in individual scans, a preliminary reduction is run without a high-pass filter, and the final image from this first reduction is then used to mask the source and to run the final reduction with high-pass filtering properly applied. 

In both pipelines the reduction process includes the following main steps: bad and saturated pixel flagging and removal, flat field correction, deglitching, high pass filtering, and map projection. Photometric maps were projected onto the final image with pixel scale 2 arcsec/pixel in the 70 $\rm \mu m$ and 100 $\mu \rm{m}$ bands and with pixel scale 3 arcsec/pixel in the 160 $\mu \rm{m}$. We also produced final maps with the native pixel scale of the detector, which were used to perform the error calculation (3.2 arcsec/pixel for the 70 $\rm \mu m$ and 100 $\mu \rm{m}$ bands and 6.4 for the 160 $\mu \rm{m}$ band). When several images at the same wavelength were available for a single target, we combined all of them to improve the signal-to-noise ratio (S/N), averaging for each pixel and using the average sigma clipping algorithm to exclude bad pixels. 

Aperture photometry was measured in final, combined images using an aperture of 6$\arcsec$ for the 70 and 100 $\mu \rm{m}$ bands and 12 $\arcsec$ for the 160 $\mu \rm{m}$ band. The annulus for sky subtraction was placed at 25-35$\rm \arcsec$ from the star. We then applied the proper aperture correction for each band\footnote{http://herschel.esac.esa.int/twiki/pub/Public/PacsCalibrationWeb/}. The photometric flux densities are listed in Table \ref{HSOphot}.

To compute photometric errors, we used native maps to minimise the impact of correlated noise, with pixel scale 3.2 arcsec/pixel in the 70 and 100 $\mu \rm{m}$ bands and 6.4 arcsec/pixel in the 160 $\mu \rm{m}$ band. Noise errors consist on the standard deviation of the photometry obtained at several sky positions surrounding the target. PACS calibration errors are 2.64$\%$, 2.75$\%$ and 4.15$\%$ for the 70 $\rm \mu m$, 100 $\rm \mu m$ and 160 $\mu \rm{m}$ bands, respectively\footnote{PICC-ME-TN-037}. Noise errors and calibration errors were added quadratically. The impact of correlated noise is negligible: final photometric uncertainties are dominated by the calibration errors, with the only exception of \object{TWA~07}. 

For non-detected sources, the derivation of upper limits is as follows; we compute the standard deviation in the sky background in several pointing surrounding the nominal position of the star on the detector; then, we compute the average value, and we use it as the sky background value. This value is multiplied by the square root of the number of pixels inside the aperture. Finally, we apply the proper aperture correction. The upper limits included in Table \ref{HSOphot} are 3-$\sigma$.

\begin{table}
\caption{Observation log}             
\label{ObsLog}      
\centering          
\begin{tabular}{lllllllll}     
\hline\hline       
Name & Obs. ID & Obs. Mode\\ 
\hline                    
\object{TWA~01} &   1342187342 & ScanMap 70/160  \\
\object{TWA~02} &   1342189163 & ScanMap 70/160  \\
\object{TWA~02} &   1342211995 & ScanMap 70/160 \\
\object{TWA~02}&   1342211996 & ScanMap 100/160  \\
\object{TWA~03A} &   1342211991 & ScanMap 70/160  \\
\object{TWA~03A} &   1342211992 & ScanMap 100/160  \\
\object{TWA~04B} &   1342188473 & ScanMap 70/160  \\
\object{TWA~04B} &   1342212634 & ScanMap 70/160  \\
\object{TWA~04B} &   1342212635 & ScanMap 100/160  \\
\object{TWA~05AB} &   1342213111 & ScanMap 70/160   \\
\object{TWA~05AB} &   1342213112 & ScanMap 100/160   \\
\object{TWA~07} &   1342188515 & ScanMap 70/160  \\
\object{TWA~07} &   1342211993 & ScanMap 70/160  \\
\object{TWA~07} &   1342211994 & ScanMap 100/160  \\
\object{TWA~10} &   1342188518 & ScanMap 70/160  \\
\object{TWA~10} &   1342213854 & ScanMap 70/160  \\
\object{TWA~10} &   1342213855 & ScanMap 100/160\\
\object{TWA~11} &   1342188519 & ScanMap 70/160 \\
\object{TWA~11} &   1342213852 & ScanMap 70/160 \\
\object{TWA~11} &   1342213853 & ScanMap 100/160\\
\object{TWA~13} &   1342213113 & ScanMap 70/160 \\
\object{TWA~13} &   1342213114 & ScanMap 100/160 \\
\object{TWA~16} &   1342188854 & ScanMap 70/160  \\
\object{TWA~16} &   1342213856 & ScanMap 70/160  \\
\object{TWA~16} &   1342213857 & ScanMap 100/160  \\
\object{TWA~16} &   1342213858 & ScanMap 70/160  \\
\object{TWA~16} &   1342213859 & ScanMap 100/160  \\
\object{TWA~21} &   1342188514 & ScanMap 70/160  \\
\object{TWA~21} &   1342211983 & ScanMAp 70/160  \\
\object{TWA~21} &   1342211984 & ScanMap 100/160  \\
\object{TWA~23} &   1342188516 & ScanMap 70/160  \\
\object{TWA~25} &   1342188517 & ScanMap 70/160  \\
\object{TWA~25} &   1342213624 & ScanMap 70/160  \\
\object{TWA~25} &   1342213625 & ScanMap 100/160  \\
\object{TWA~25} &   1342213626 & ScanMap 70/160  \\
\object{TWA~25} &   1342213627 & ScanMap 100/160  \\
\hline                  
\end{tabular}
\end{table}

\subsection{Spectroscopic data reduction}
Nine TWA stars were observed with PACS in LineScan spectroscopic mode. The line targeted with the LineScan observations was [OI] at 63.18 $\mu \rm{m}$. PACS spectra were reduced using HIPE 8 with the latest version of the pipeline and the proper calibration files. 

Spectra were extracted from the central spaxel when the observations were well centred there. When the observations were not properly centred on the source, we extracted the spectrum from the spaxel with the highest continuum signal. The extracted spectrum was then aperture corrected to account for flux loss in the surrounding spaxels. The line spectra from PACS typically show higher noise levels near the spectrum edges. To account for that effect, we exclude from the 63 $\mu \rm{m}$ spectra any wavelength shorter than 63.0 $\mu \rm{m}$ or longer than 63.4 $\mu \rm{m}$.

Line fluxes were computed by applying a Gaussian fit to the line profile and calculating the integrated flux from that fit. Upper limits were computed by integrating a gaussian with a width equal to the instrumental full width half maximum (FWHM) at the central wavelength, and maximum equal to three times the standard deviation of the continuum. Therefore, computed upper limits are 3-$\sigma$.  Line fluxes are shown in Table \ref{HSOspec}.

\section{Results}

\subsection{Herschel-PACS photometry} 
Within the 12 targets observed at 70 $\mu \rm{m}$, we detected four sources. At 100 $\mu \rm{m}$, we have detected four systems out of nine observed. All the targets detected in the 70 $\mu \rm{m}$ and 100 $\mu \rm{m}$ bands are also detected at 160 $\mu \rm{m}$, and all of them show emission in excess above the photosphere, in agreement with the presence of circumstellar material. Excess fractions are $\rm 0.33^{+0.15}_{-0.10}$, $\rm 0.44^{+0.16}_{-0.14}$ and $\rm 0.36^{+0.14}_{-0.10}$ at 70 $\mu \rm{m}$, 100 $\mu \rm{m}$ and 160 $\mu \rm{m}$, respectively \citep[see][for a detailed description of the method used to compute errors on the disc fractions]{Burgasser2003}. We have detected for the first time the 100 $\mu \rm{m}$ and 160 $\mu \rm{m}$ emission toward \object{TWA~07}, and 100 $\mu \rm{m}$ emission toward \object{TWA~03A}.

\object{TWA~13A} was considered a far-IR excess source by \cite{Low2005}. They found a 27.6 mJy flux density at 70 $\mu \rm{m}$. But \cite{Plavchan2009} showed that the 70 $\mu \rm{m}$ emission is coincident with a background galaxy and not with \object{TWA~13A}. Our combined images for \object{TWA~13A} do not show any source at the nominal position of the star, therefore we agree with \cite{Plavchan2009} that there is no excess associated with \object{TWA~13A}. However there is a background galaxy that could have polluted the \cite{Low2005} observation at RA=11:21:16.6, DEC=-34:46:38.6, with a flux density of 70 mJy at 100 $\mu \rm{m}$ and 93 mJy at  160 $\mu \rm{m}$. 

\begin{table}
\caption{TWA PACS photometry}             
\label{HSOphot}      
\centering          
\begin{tabular}{c c c c }     
\hline\hline       
name & 70 $\mu \rm{m}$  & 100 $\mu \rm{m}$ &  160 $\mu \rm{m}$  \\ 
\hline
	& (mJy)  & (mJy)  & (mJy) \\ 
\hline                    
\object{TWA~01}$\rm ^{1}$  & 3900 $\pm$ 100 & -- & 7380 $\pm$ 300  \\
\object{TWA~03A} & --  & 650 $\pm$ 19 & 459 $\pm$ 19  \\
\object{TWA~04B} & 6241 $\pm$ 165 & 4269 $\pm$ 117 & 2382 $\pm$ 100  \\
\object{TWA~07}  & 77 $\pm$ 7 & 58 $\pm$ 4 & 42 $\pm$ 9   \\
\object{TWA~11} & 4980 $\pm$ 131 & 3553 $\pm$ 97 & 1653 $\pm$ 68  \\
\hline
\object{TWA~02}  & $<$ 11.0  & $<$ 6.4 & $<$ 19 \\
\object{TWA~05} & $<$ 2.2  & --  & $<$  11.0  \\
\object{TWA~10}  & $<$ 4.1 & -- & $<$ 12.0  \\
\object{TWA~12}  & $<$ 2.8  & -- & $<$ 9.0 \\
\object{TWA~13A} & -- &  $<$ 6.4 & $<$ 19.0  \\
\object{TWA~16}	& $<$ 4.4 & $<$ 4.6 & $<$ 8.6 \\
\object{TWA~21}  & $<$ 4.0 &  -- & $<$ 14.0   \\
\object{TWA~23}  & $<$ 5.1 & $<$ 6.0 & $<$ 13.0  \\
\object{TWA~25}  & $<$ 5.7 & $<$ 6.3  & $<$ 12.0   \\
\hline                  
\end{tabular}
\tablefoot{Errors include a calibration error of 2.64$\rm \%$, 2.75$\rm \%$ and 4.15 $\rm \%$ for the 70 $\mu \rm{m}$, 100 $\mu \rm{m}$, and 160 $\mu \rm{m}$ bands respectively (see text). Upper limits are 3-$\sigma$ (1): from \cite{Thi2010}. }
\end{table}

\begin{figure}[!t]
\begin{center}
   \centering
     \includegraphics[scale=0.5]{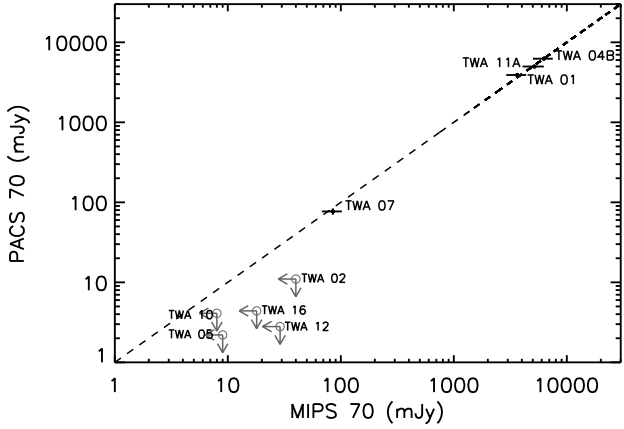}
   \caption{Comparison plot for PACS 70 $\mu \rm{m}$ flux densities vs MIPS 70 $\mu \rm{m}$ flux densities. The black dashed line represents the one to one relation. Displayed upper limits are 3-$\sigma$.}
   \label{PACS_MIPS_COMP}
\end{center}
\end{figure}

In Fig. \ref{PACS_MIPS_COMP}, we show a comparison of the photometry at 70 $\mu \rm{m}$ from MIPS with that from PACS for the remaining sources. We see that the agreement between MIPS and PACS for detected objects is good; both datasets are in agreement within the calibration uncertainties, and sources are therefore not variable. Any discrepancy can be due to problems in the calibration of the instruments, to the larger beam size in MIPS or to real variability. The most probable explanation lies on the larger beam size used by MIPS, which can include some contamination from the background. The upper limits from PACS are pushing further down to photospheric values, and are typically $\rm \sim$5 times smaller than previous upper limits from MIPS.

\begin{figure}[!h]
\begin{center}
   \centering
     \includegraphics[scale=0.45]{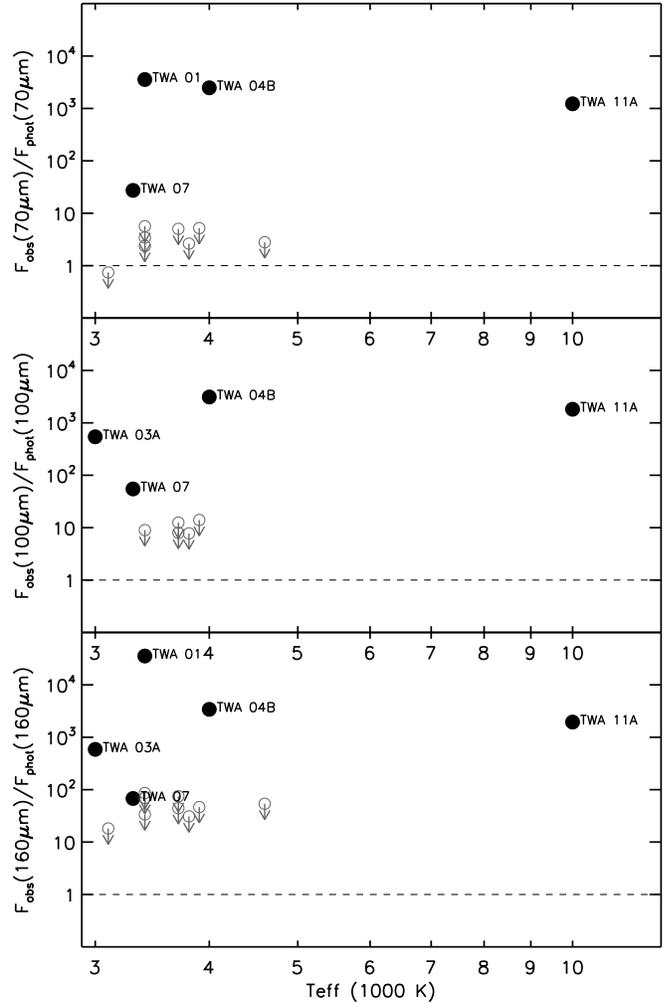}
     \caption{Excesses at 70 (top), 100 (middle) and 160 (bottom) $\mu \rm{m}$ band versus $\rm T_{eff}$. Solid dots represent true detections, while empty circles with arrows represent upper limits. The black horizontal dashed line shows the limit for photospheric emission. Displayed upper limits are 3-$\sigma$.}
   \label{ExcessVSTeff}
\end{center}
\end{figure}

In Fig. \ref{ExcessVSTeff}, we plot the 70 $\mu \rm{m}$, 100 $\mu \rm{m}$ and 160 $\mu \rm{m}$ excess versus the effective temperature of the central star. The excesses are computed as the observed flux density over the photospheric flux density at each band. Photospheric flux densities are extracted from the models in Sec. \ref{StParam}. The behaviour is the same for every PACS band; a system bright at 70 $\mu \rm{m}$ is also bright at 100 and 160 $\mu \rm{m}$. The bimodal distribution of excesses seen at 24 $\mu \rm{m}$ and lower mid-IR wavelengths \citep{Weinberger2004,Low2005} is not as pronounced at 70 $\mu \rm{m}$, but we note that the \object{TWA~07} excess at 70 $\mu \rm{m}$ is more than one order of magnitude smaller than any other detected excess in TWA, and that upper limits at 70 $\mu \rm{m}$ are near photospheric values, so probably no disc emission is  present at this wavelength for those undetected sources. The bimodal nature of the distribution vanishes at 100 and 160 $\mu \rm{m}$; excesses at 100 and 160 $\mu \rm{m}$ cover three orders of magnitude, with ratios of the order of 10, 100 and 1000 depending on the star. Upper limits are far from the photospheric value at 100 $\mu \rm{m}$, where typical values are 10 times larger than the photosphere. Finally upper limits at 160 $\mu \rm{m}$ are 10 to 100 times larger than the photospheric flux densities. Therefore, we can not exclude the presence of faint, cold discs in those targets.

In order to check for spatially resolved sources, we calculated azimuthally averaged radial profiles in all three PACS bands for every detected object and compared the results with the azimuthally averaged radial profile of a model point spread function (PSF) from the calibration star \object{$\alpha$ Boo}. We obtained the radial profiles making use of the IRAF task \textit{pradprof}.  Within the errors, none of the sources is extended in any of the PACS bands.

\subsection{Herschel-PACS spectroscopy}
Among the nine TWA members observed with PACS in spectral line mode, we only detected the continuum level at 63 and 189  $\mu \rm{m}$ in four sources, namely \object{TWA~01}, \object{TWA 03A}, \object{TWA~04B} and \object{TWA~11A}, i. e. , those objects detected also with PACS photometry with the exception of \object{TWA~07}, whose spectroscopic continuum level is below the detection limit. 

\begin{figure*}[!ht]
\begin{center}
   \centering
     \includegraphics[scale=0.33]{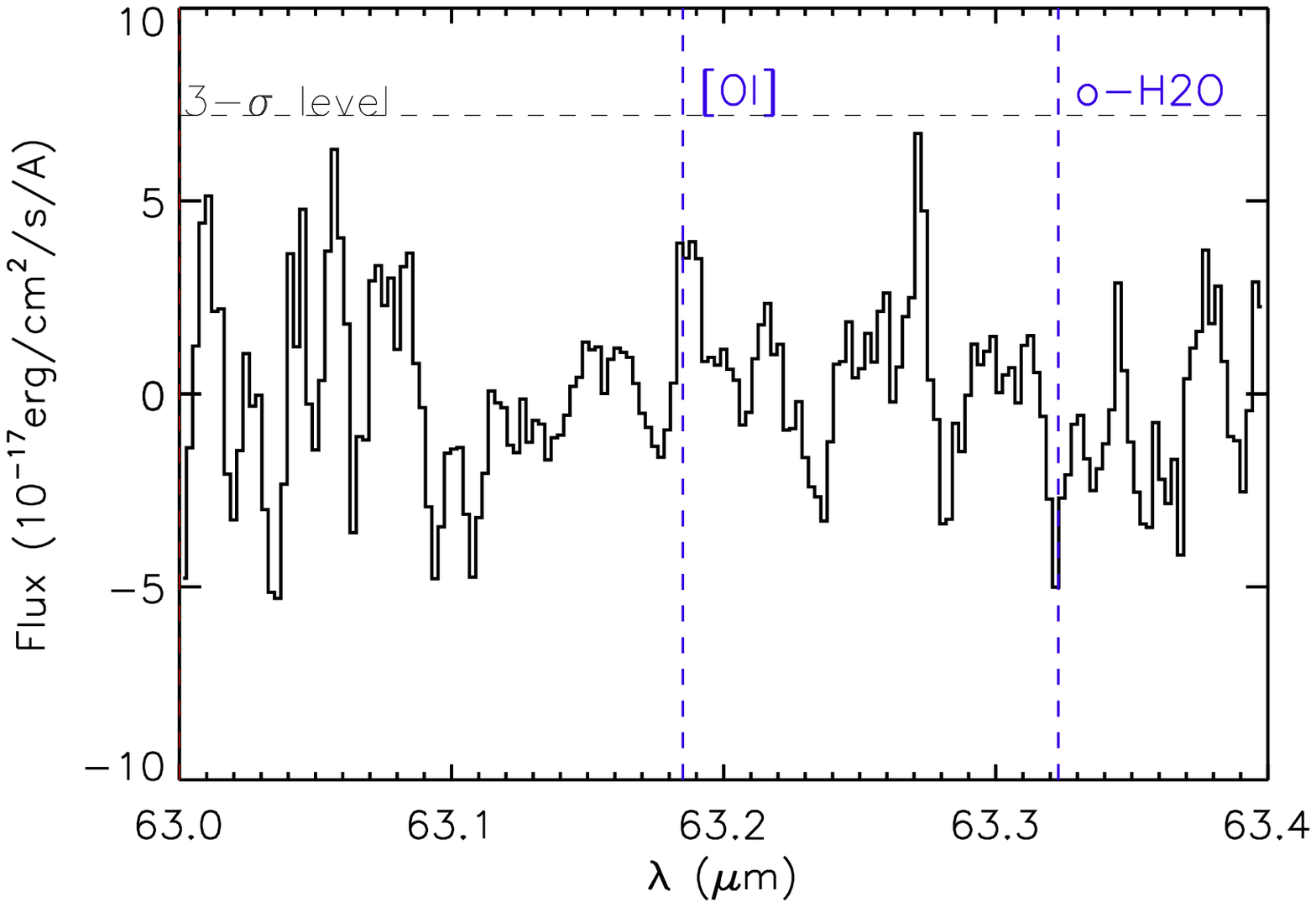}\includegraphics[scale=0.33]{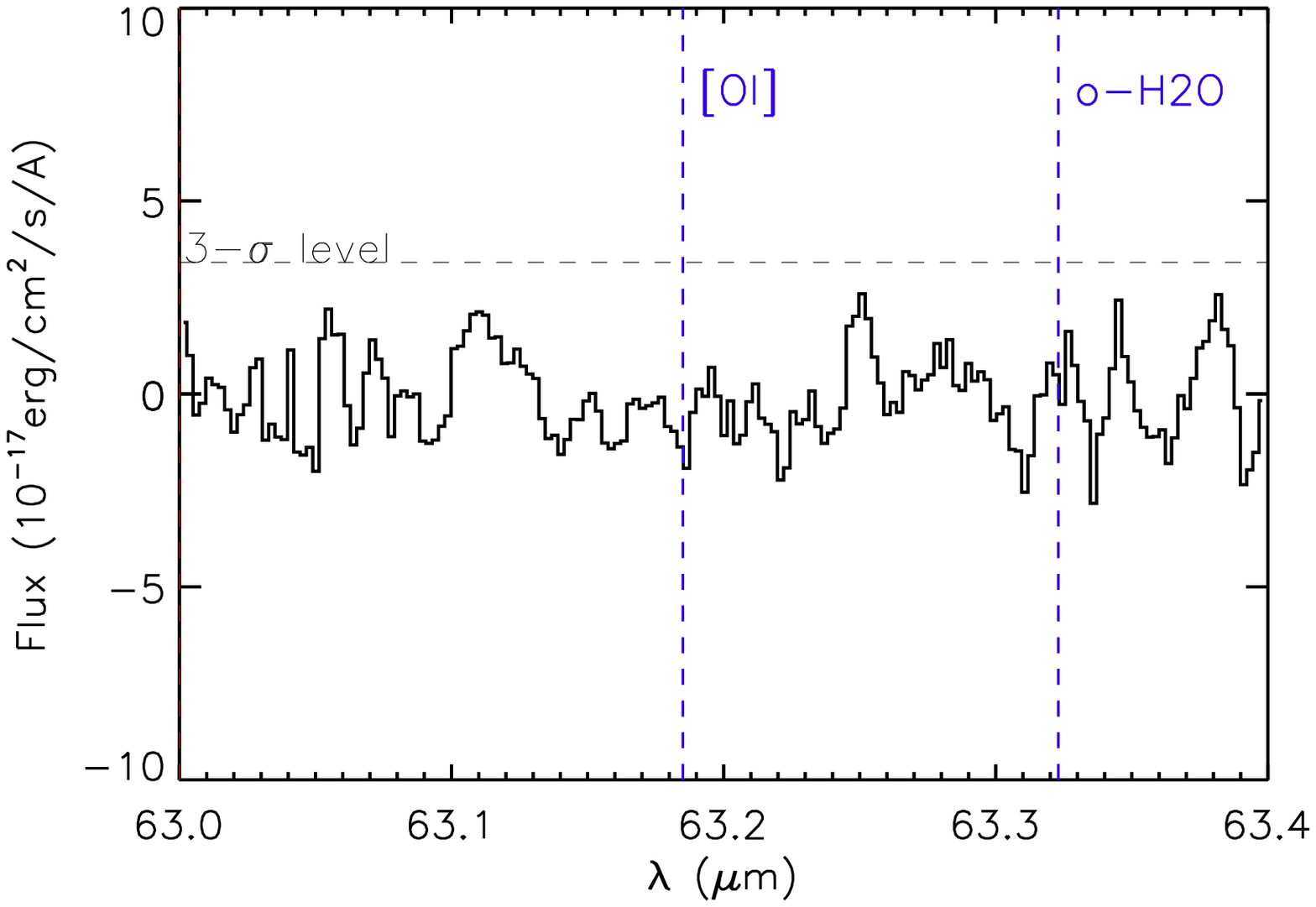}\includegraphics[scale=0.33]{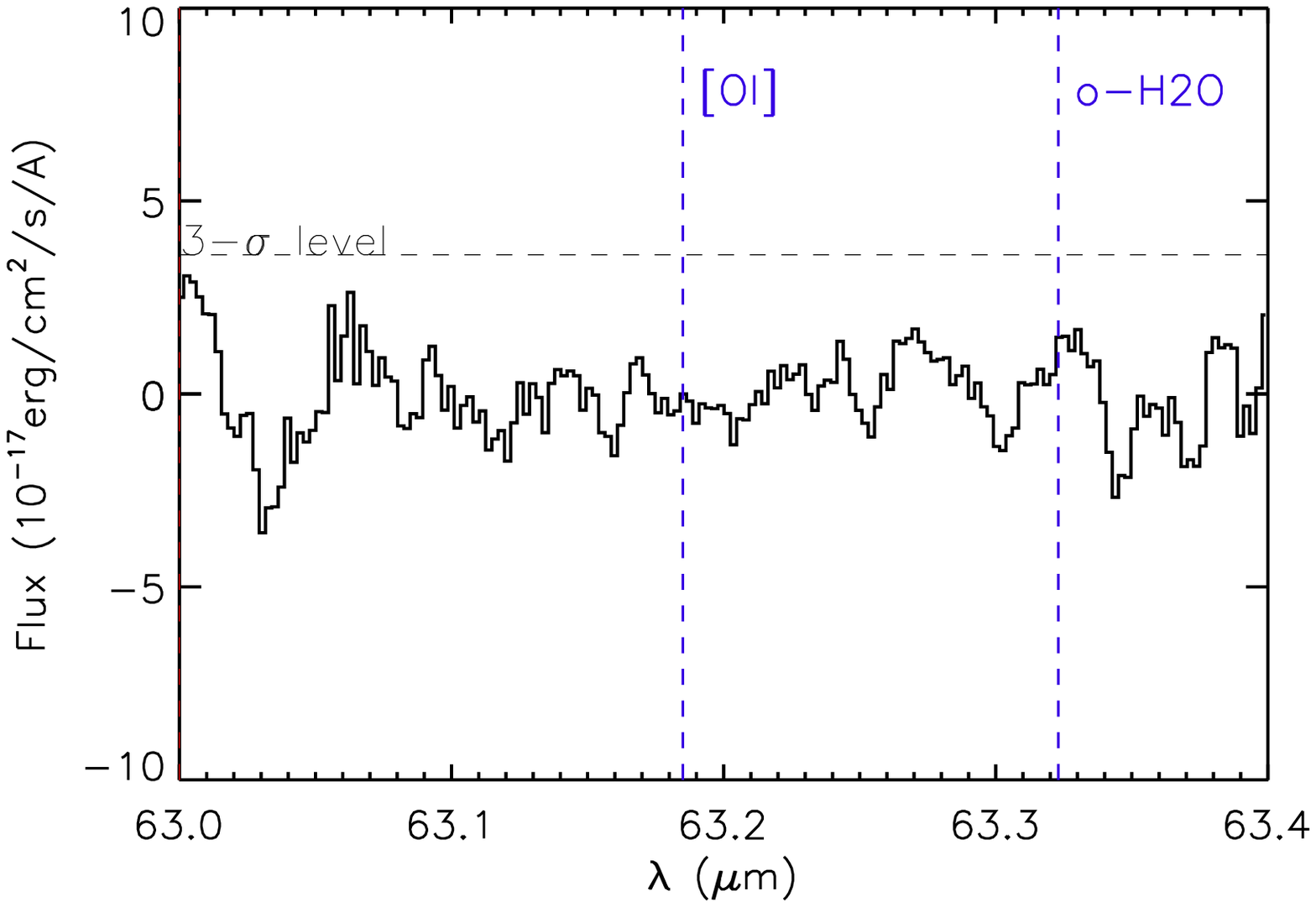}
   \caption{Continuum subtracted spectra of TWA members showing IR excess but no line emission at 63 $\mu \rm{m}$. The vertical blue dashed lines represent the position of the [OI] and o-$\rm H_{2}O$. From left to right, targets are: TWA 03A, TWA 07 and TWA 11A. We show 3-$\sigma$ limits as horizontal, black dashed lines.}
   \label{LineSpecND}
\end{center}
\end{figure*}

\begin{figure}[!ht]
\begin{center}
   \centering
     \includegraphics[scale=0.5]{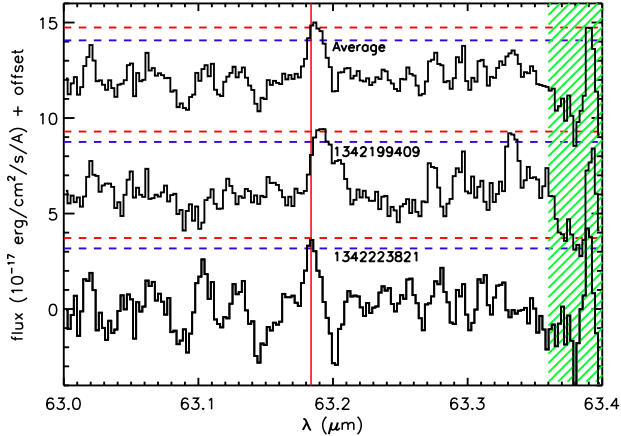}
   \caption{The three different continuum subtracted spectra in the region around 63 $\mu \rm{m}$ for \object{TWA~04B}. An arbitrary vertical shift was applied to the spectra for easy comparison. The vertical red line shows the rest wavelength of the [OI] 63.18 $\mu \rm{m}$ emission line. The green dashed area shows the position of a region where the noise dominates the emission. The red dashed horizontal lines show the the 3-$\sigma$ limit for the three spectra. The blue dashed horizontal lines show the 3-$\sigma$ limit when points inside the green dashed area are excluded.}
   \label{TWA04Bspec}
\end{center}
\end{figure}

Only two TWA members show [OI] emission at 63.18 $\mu \rm{m}$; \object{TWA~01} and \object{TWA~04B}. Line fluxes are given in Table \ref{HSOspec}. We note that [OI] line emission in \object{TWA~01} was studied and modelled by \cite{Thi2010} in detail. Continuum subtracted spectra for TWA members showing IR excess with no line detections at 63 $\mu \rm{m}$ are shown in Fig. \ref{LineSpecND}. To compute lines fluxes and upper limits, we fitted a second order polynomial to the baseline to subtract the continuum.

TWA 04B was observed at two different epochs. The resulting continuum subtracted spectra are shown in Fig. \ref{TWA04Bspec}. The first one (OBSID 1342199409) shows a clear 3-$\sigma$ detection, irrespective of the region used to compute the continuum. The second one is not as robust;  it is a 3-$\sigma$ detection if we exclude the noisy region that goes from 63.36 to 63.40 $\mu \rm{m}$ (shaded in green in Fig. \ref{TWA04Bspec}) from the continuum computation. But if we include this region, it is at the limit of a 3-$\sigma$ detection. The average spectrum shows a 3-$\sigma$ detection independently of the region used to derive the continuum, so we consider that this is a real detection. Due to a shift in the line center in both observations with respect to the rest frame wavelength of [OI] (63.185 $\rm \mu m$), the combination of the two epochs does not result in the expected increase in S/N. The [OI] detection in \object{TWA~04B} will be discussed further in Sec. \ref{TWA04B}.

\begin{table}[!hd]
\caption{TWA PACS line fluxes}             
\label{HSOspec}              
\begin{tabular}{llll}     
\hline\hline       
name & $\rm{ [OI]}_{\rm{63.18} \mu \rm{m}}$& $\rm{ o-H_{2}O}_{\rm{63.32} \mu \rm{m}}$ & $\rm{ S/N}_{\rm{63} \mu \rm{m}}$ \\ 
\hline
--	& $\rm (10^{-18} W/m^{2})$ & $\rm (10^{-18} W/m^{2})$ & --  \\ 
\hline     
\object{TWA~01} & $\rm 36.5 \pm 12.1 $  & $\rm < 2.42$ & 13.3 \\               
\object{TWA~02} & $\rm < 8.3 $ & $\rm < 8.3 $ & 0.19  \\
\object{TWA~03A} &  $\rm < 5.8 $ & $\rm < 5.8 $ & 33.9 \\
\object{TWA~04B} $\rm ^{1}$ & $\rm 6.4 \pm 1.5 $ & $\rm < 1.74$ & 41.8  \\
\object{TWA~04B} $\rm ^{2}$ & $\rm 4.0 \pm 1.0 $ & $\rm < 1.34$ & 46.3  \\ 
\object{TWA~04B} $\rm ^{3}$ & $\rm 4.1 \pm 1.2 $ & $\rm < 1.24$ & 56.9 \\
\object{TWA~07} & $\rm < 4.6 $ & $\rm < 4.6 $ & 1.04  \\
\object{TWA~10} & $\rm < 4.5 $ & $\rm < 4.5 $ & 0.35  \\
\object{TWA~11A} & $\rm < 4.7$ & $\rm < 4.7$& 39.67 \\
\object{TWA~13A} & $\rm < 4.5$ & $\rm < 4.5$ & 0.32 \\ 
\object{TWA~23} & $\rm < 5.4 $ & $\rm < 5.4 $ & 0.28 \\ 
\hline                  
\end{tabular}
\tablefoot{Listed values for S/N are continuum S/N. (1): OBSID 1342199409. (2): OBSID 1342223821. (3): Averaged spectrum.}
\end{table}

\section{Models of dust discs}
\begin{figure*}[!ht]
\begin{center}
   \centering
     \includegraphics[scale=0.5]{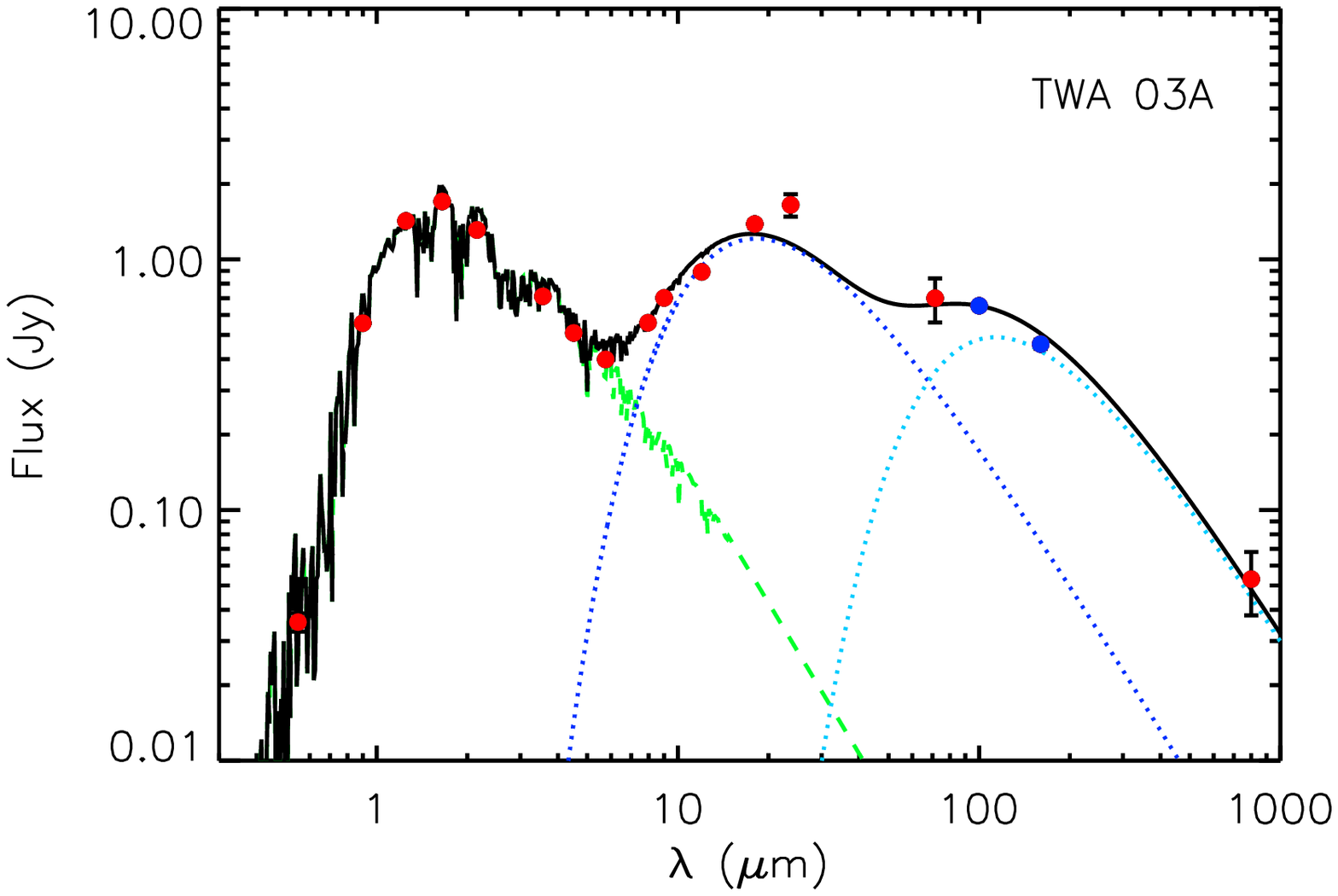}
     \includegraphics[scale=0.5]{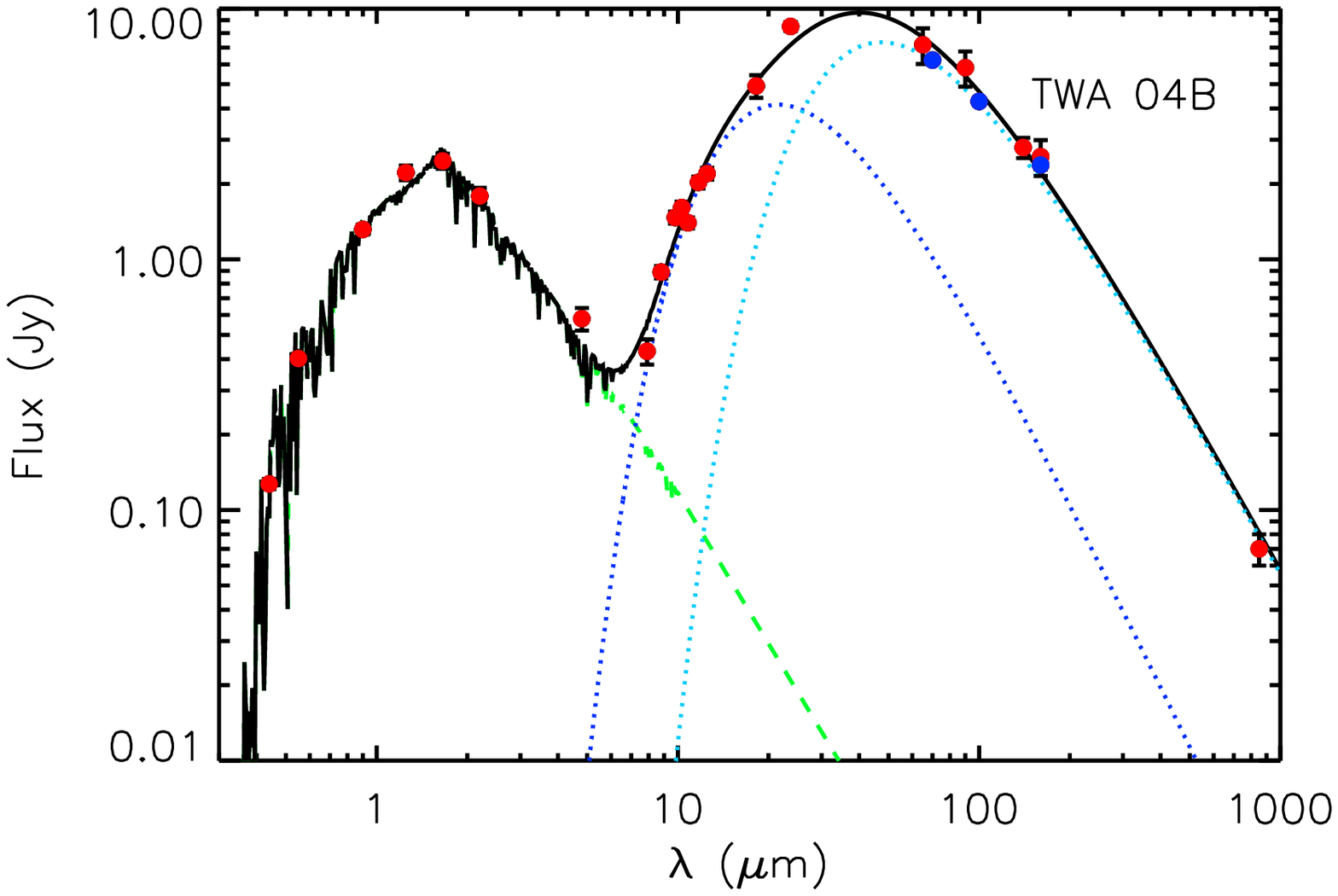}
     \includegraphics[scale=0.5]{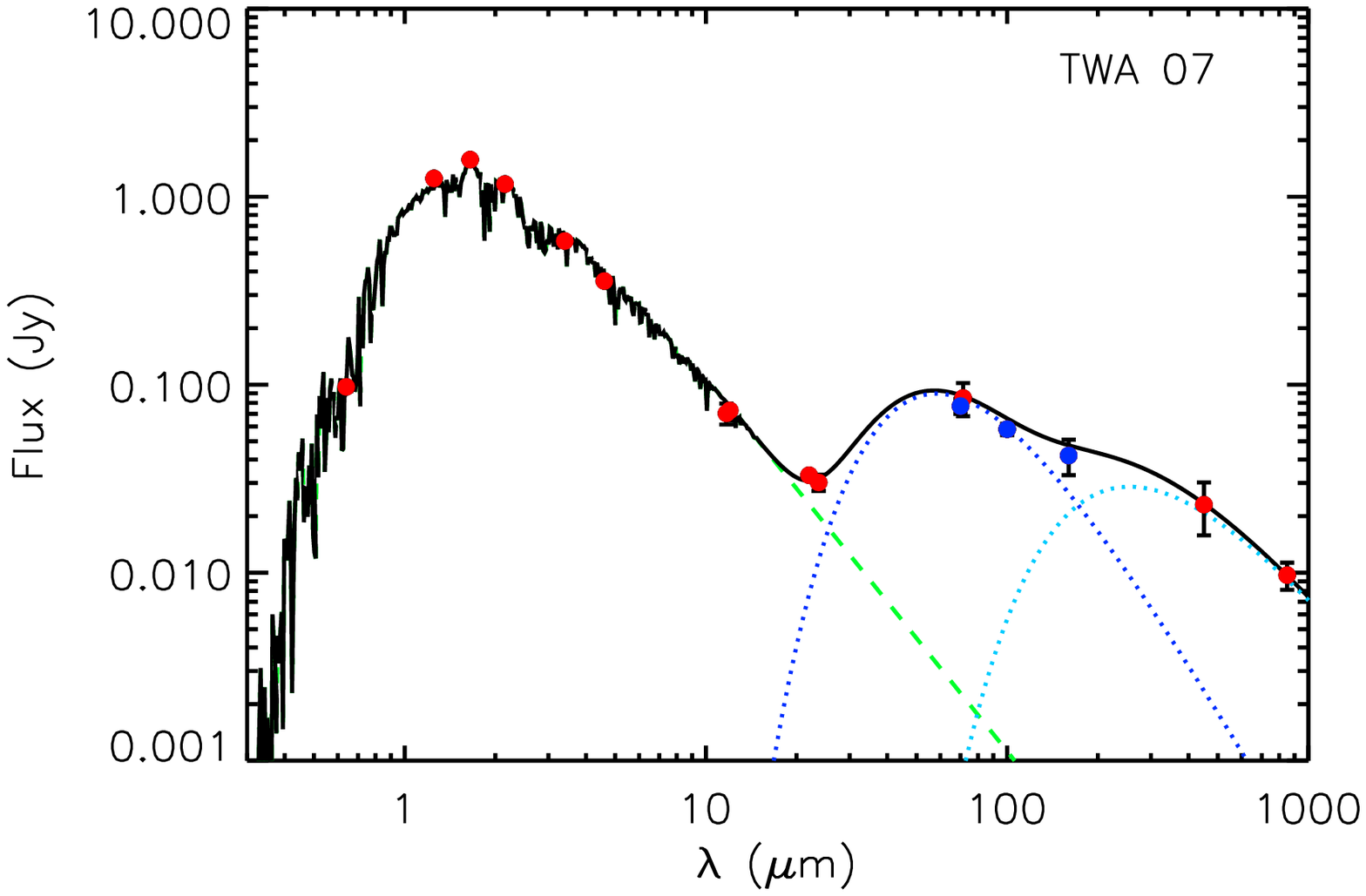}
     \includegraphics[scale=0.5]{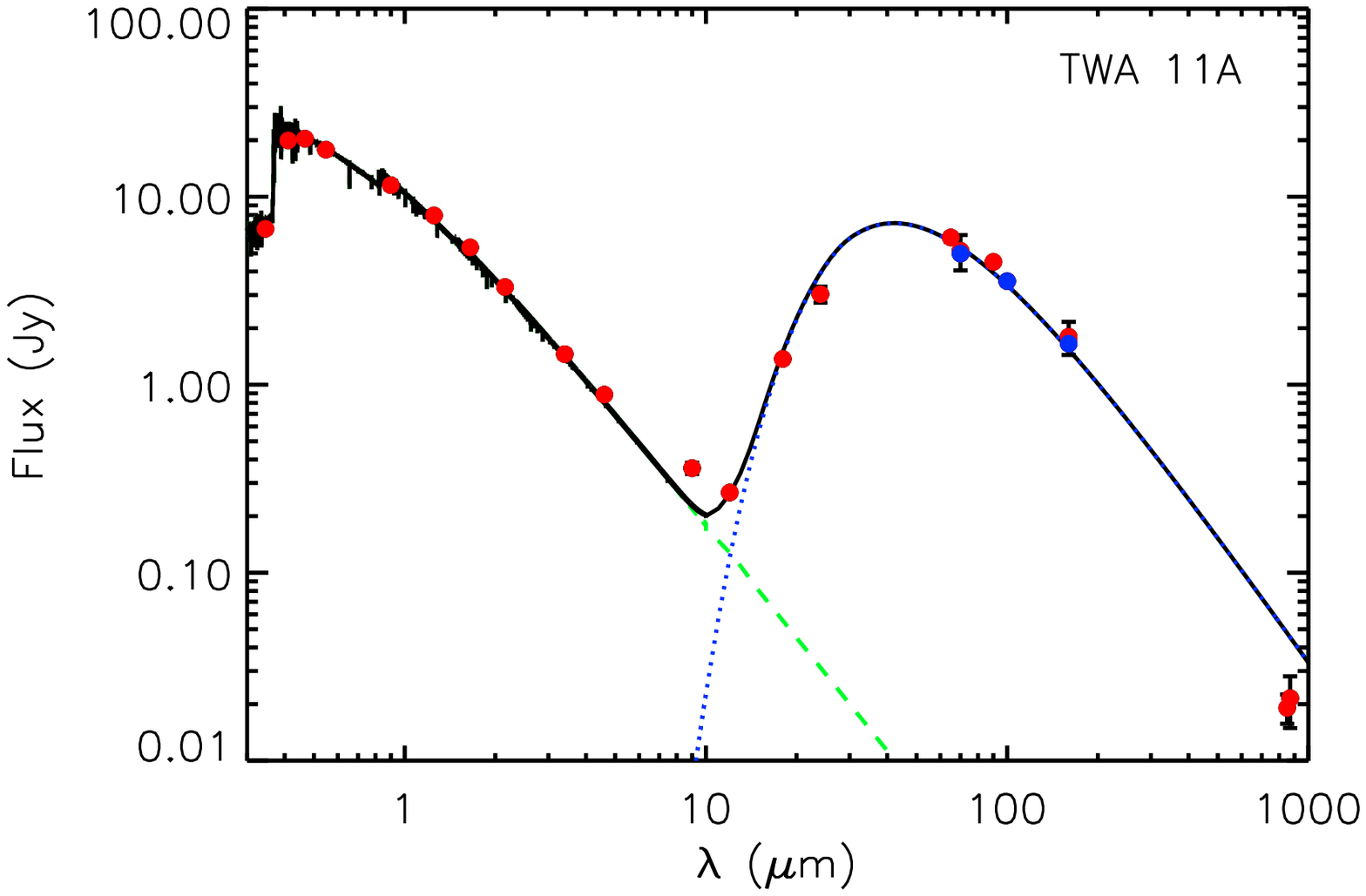}
     \caption{Blackbody models for TWA members. Dark and light blue dashed lines depict the different blackbody components for the two blackbodies model. The blue dots represent PACS observations at 70, 100 and 160 $\mu \rm{m}$. The green line represents the photospheric contribution.}
   \label{BBmods}
\end{center}
\end{figure*}

For every star in the sample, we built the SED including the new \textit{Herschel}-PACS data and aiming to compare the observed data with modified blackbody models to estimate basic dust properties in the TWA association. We are aware that this approach may be very simplistic for some of these targets, but it is a good starting point for estimating dust temperatures. In any case, each individual object is discussed 
in the next sections. Figure \ref{BBmods} includes SEDs for TWA members detected with \textit{Herschel}. The disc models were built using the NextGen models from Sec. \ref{StParam} plus a modified blackbody to describe the dust emission, defined as

\begin{equation}
F_{continuum}=B_{\nu}(T_{dust}) \times (\lambda_{0}/\lambda)^{\beta}
\end{equation}
with $\lambda_{0}\rm{=2} \pi a_{min}$ and $\beta$=0 for $\lambda < \lambda_{0}$, where ${\rm B}_{\nu}$ is the Planck function. The choice for $\rm a_{min}$ depends on the target. \object{TWA~11A} is an A0 star, and therefore grains below the blow-out size ($\rm a_{blow}$) are expected to be blown out from the system 
on timescales that are much shorter than an orbital period. Therefore, for \object{TWA~11A} $\rm a_{min}=a_{blow}$, which can be computed using	

\begin{equation}
{a_{blow} \over 1 \mu m} = 1.15 {L_{*} \over L_{\odot}} {M_{\odot} \over M_{*}} {1 g cm^{-3} \over \rho}
\end{equation}
\citep{Backman1993} with $\rho \rm{=2.5 ~g/cm^{3}}$, which is valid for astrosilicates. We get $\rm a_{blow}=28 \mu m$. On the other hand, \object{TWA~01}, \object{TWA~03A}, \object{TWA~04B} and \object{TWA~07} are M stars, where no grains are expected to be be blown out from the system as a result of radiation pressure. For these sources, we used $\rm a_{min}=0.1 \mu m$ following the models from \cite{Augereau2006} for \object{AU~Mic}.

Both $\beta$  and $\rm T_{dust}$ are free parameters, and $\beta$ can take any value from 0 to 2. A value of 2 was found for unprocessed interstellar grains \citep{Hildebrand1983}, while a value of 0 indicates dust grains radiating as pure blackbodies. \cite{Miyake1993} demonstrated that grain growth results in $\beta$ values that are significantly lower than the standard interstellar medium (ISM) value of 2 and that grain distributions with $\rm{ a_{max} \gtrsim 1000} $ $\mu \rm{m}$  result in $\beta \rm{< 1}$. Therefore, the $\beta$ value from the modified blackbody fit gives us some information about the grain size distribution in TWA discs.

The best model was determined through $\chi^{\rm{2}}$ minimisation, with
\begin{equation}
 \chi^{2}={1 \over \nu} \sum_{i=1}^{N} {(F_{obs,i}-F_{mod,i})^{2} \over \sigma_{i}^{2} }
\end{equation}
where $N$ is the total number of photometric points used in the fitting, $\nu = N-n$ the degrees of freedom, n the number of free parameters, $F_{obs,i}$ the observed flux density of the \textit{i}-th photometric point, $F_{mod,i}$ the  \textit{i}-th model flux density, and $\sigma_{i}$ is the error of the \textit{i}-th photometric point. We used a genetic algorithm to search the models that minimise $\chi^{2}$. Basically, a range (minimum and maximum) for each parameter must be supplied to the algorithm, which follows these steps: 1) generate a random population of  N individuals (i.e, sets of parameters, with N being user-defined and usually between 10 to 40) within the range limits for every parameter to fit; 2) generate the models corresponding to the random individuals; 3) select the best 10$\rm \%$ of individuals based on a $\chi^{\rm{2}}$ minimisation, and 4) use this selection of best models to generate the next population of individuals with values in the range -20$\rm \%$ to +20$\rm \%$  of the model parameters from step 3. The process is iteratively repeated until a stable point is reached, i. e., when the best $\chi^2$ does not change more than 1\% with respect to the best value on the previous generation. At this point, a random population of $\rm 10 \times N$ new individuals is added to destabilise the achieved minimum owing to avoid local minima.  This process is then repeated ten times. Hence, the final value must remain stable during more than 100 generations with ten inclusions of $\rm 10 \times N$ random points in the whole parameter space. Moreover, the random population of $\rm 10 \times N$ created every time a stable point is reached is used to compute the final errors in the parameters and to check for possible multi-valued solutions. In particular, we measure the 90\% countours with respect to the best $\chi^2$ value obtained by the genetic algorithm.

From these fits, we get estimates of the dust temperature and IR excess. This analysis is useful since it relies on very few assumptions and allows easy comparison with other studies. For the sources with no IR excess detected we compute upper limits for the IR excesses by considering the flux density upper limits as detections.
For \object{TWA~01}, \object{TWA~03A}, \object{TWA~04B} and \object{TWA~07}, the SED is reproduced better using two blackbodies instead of only one. That we need two blackbodies to reproduce the emission could be pointing to the presence of two populations of grains at different radii or could be the signpost of a complex system. For \object{TWA~11A}, the inclusion of a second blackbody does not result in a big improvement. Results of the fitting process are listed in Table \ref{BBtable}, and model SEDs can be found in Fig. \ref{BBmods}. 

Using the dust temperature from the blackbody models, we can estimate the inner radius of the disc using  
\begin{equation}
R_{\rm in} > \frac{1}{2} R_{*}\left(T_{*} \over T_{\rm
  dust}\right)^{(4+ \beta)/2}
\end{equation}
\citep{Beckwith1990}. Furthermore, we can estimate the dust mass in the disc by using the following expression, which is valid for optically thin discs:
\begin{equation}\label{eq:dustMass}
M_{dust}={F_{\nu}(\lambda_{0})D^{2} \over \kappa_{\nu} B_{\nu}(T_{dust})}
\end{equation}
where $D$ is the
distance to the star, $\rm{ B}_{\nu}(\rm{T_{dust}})$ can be approximated by
the Rayleigh-Jeans regime, $\kappa_{\nu}\rm{=2\times (1.3 mm /}
\lambda)^{\beta}~\rm{cm^{2}g^{-1}}$ \citep{Beckwith1990}, and $ \rm{F}_{\nu}(\lambda_{0})$ is the integrated flux density at a given wavelength emitting at the Rayleigh-Jeans regime. Considering that for some targets, the far-IR and sub-millimetre flux densities seem to arise from different radii, we computed dust masses using the PACS flux density  at 160 $\mu \rm{m}$ and a sub-millimetre flux density from the literature. Disc radii and dust masses are shown in Table \ref{BBtable}.

\begin{table*}[ht!]
\caption{Photosphere plus modified blackbody fitting and disc parameters}             
\label{BBtable}      
\centering          
\begin{tabular}{llllllllll}     
\hline\hline       
Name & $\rm T_{1}$ & $ \beta_{1}$ & $\rm T_{2}$ & $ \beta_{2}$ &$\rm L_{IR}/L_{*}$ & $\rm R_{in,1}$ & $\rm R_{in,2}$  & $\rm{ M}_{\rm{dust,160} \mu m}$ & $\rm M_{dust,submm}$ \\ 
\hline
 -- & (K)  & -- & (K) & -- & --  & (AU) & (AU) & ($\rm M_{\oplus}$)  \\ 
\hline           
\object{TWA~03A} & $\rm 280 \pm 5 $& $\rm 0^{+0.06}$ & $\rm 45 \pm 4 $& $\rm 0^{+0.09}$ & $\rm 9.8 \times 10^{-2}$ & $\rm 0.416 \pm 0.015$ & $\rm 16.1 \pm 5.8$ & $\rm 0.133 \pm 0.042$ & $\rm 0.384 \pm 0.159$ \\
\object{TWA~04B} & $190 \pm 10$ & $\rm 0.72 \pm 0.55$ & $\rm 102 \pm 2$ & $\rm 0.185 \pm 0.023$ &  $\rm 2.4 \times 10^{-1}$ & $\rm 4.8  \pm 1.0$ & $\rm 7.9 \pm 1.6$ & 0.259 $\pm$ 0.029 & 0.256 $\pm$ 0.064 \\
\object{TWA~07} & $\rm 66 \pm 5$ & $\rm 0.86 \pm 0.23$ & $\rm 20 \pm 2$ & $\rm 0^{+0.1}$ & $\rm 2.2 \times 10^{-3}$ & $\rm 38 \pm 10 $ & $\rm 75 \pm 15$ &  $\rm 0.022 \pm 0.011$ & $\rm 0.146 \pm 0.045$\\
\object{TWA~11A} & $\rm 108 \pm 5$ & $\rm 0.3^{+0.2}_{-0.1}$ & -- & -- &$\rm 4.6 \times 10^{-3}$ & $\rm59 \pm 19$ & -- & $\rm 0.271 \pm 0.109$ & $\rm 0.146 \pm 0.042$\\
\hline
\end{tabular}
\end{table*}

\section{Gas and dust content in TWA circumstellar discs}
We have analysed the main results from our blackbody (BB) dust models and the Herschel spectroscopic data. In the following, we discuss the individual sources.

\subsection{\object{TWA~01}}
PACS observations of \object{TWA~01} from the GASPS programme have been previously studied and modelled in detail by \cite{Thi2010},  where they modelled the circumstellar dust with a total dust mass of $\rm \sim 63~M_{\oplus}$. \cite{Bergin2013} report detecting hydrogen deuteride (HD), and derived a gas mass $\rm > 1.8 \times 10^{5}~M_{\oplus}$, high enough to form a planetary system at the age of $\rm ~10~Myr$. As shown in section \ref{Acc}, \object{TWA~01} is the strongest accretor among the sample. The blackbody model for \object{TWA~01} produces a poor fit, especially when compared with other TWA members such as \object{TWA~04B}, \object{TWA~07}, and \object{TWA~11A}. This is due to the more complex spatial distribution of dust around \object{TWA~01} and to its optical thickness. We refer the reader to the paper by \cite{Thi2010} for a more detailed analysis of this source.

\subsection{\object{TWA~03A}}
\cite{Andrews2010} performed high-resolution observations of the thermal continuum emission towards \object{TWA~03A} at 880 $\mu \rm{m}$ with SMA. By modelling the 880 $\mu \rm{m}$ flux visibility and the SED, they deduced $\rm R_{out} \sim 15-25~AU$ (the expected radius if the disc around \object{TWA~03A} is truncated by \object{TWA~03B}), $\rm R_{in} \sim 0.4~AU$, and a dust mass of $\rm 2.3$\,$\rm M_{\oplus}$. 

The best model for \object{TWA~03A} needs a population of hot grains ($\rm T=280~K$) at $\rm \sim$ 0.4 AU to reproduce the mid-IR emission and a second population of cold grains  ($\rm T=40~K$) at 16 AU. We note that the model fails to reproduce the MIPS observation at 24 $\mu \rm{m}$. The flux density at 160 $\mu \rm{m}$ is not at the Rayleigh-Jeans regime, and therefore the mass computed using this flux density is a lower limit. A more realistic value for the dust mass can be obtained if the flux density in Equation \ref{eq:dustMass} is the James Clerk Maxwell Telescope (JCMT) flux density at 800 $\mu \rm{m}$ from \cite{Jensen1996}. We get $\rm \sim 0.38~M_{\oplus}$, about six lower smaller than the value by \cite{Andrews2010}.

\cite{Muzerolle2000} studied $\rm H_{\alpha}$ emission in \object{TWA~03A} and concluded that there is ongoing gas accretion in the system about one order of magnitude larger than the average value for discs in Taurus. \cite{Huenemoerder2007} studied the soft X-ray emission from the star and concluded that the shape of the spectrum was attributable to accretion shocks. The compilation of \object{TWA~03A} $\rm{H}_{\alpha}$ equivalent widths by \cite{Barrado2006} shown in Fig. \ref{AccretionPlot} also agree with ongoing accretion as pointed out in section \ref{Acc}. This contrasts with the non-detection of [OI] emission at 63 $\mu \rm{m}$ and is discussed in Sec. \ref{TWAgas_sec}.

\subsection{\object{TWA~04B}}\label{TWA04B}
\object{TWA~04B} is part of a peculiar quadruple system, which is two spectroscopic binaries A and B orbiting each other. The system shows a prominent infrared excess attributed to a circumbinary disc around the B pair \citep{Koerner2000, Prato2001,Furlan2007}, where both stars are separated by $\rm \sim$1 AU. \cite{Skinner1992} reported the detection of a silicate emission feature around 10 $\mu \rm{m}$, and indicated the presence of grains as small as 0.01 $\mu \rm{m}$. 

Different studies have modelled the infrared emission in TWA 04B in the past, showing inner radii ranging from 2 AU to 5 AU and outer radii in the range from 10 to 18 AU. Some of these studies modelled the SED of \object{TWA~04B} using a complex geometry with two discs \citep{Akeson2007, Furlan2007, Andrews2010}. The most recent work by \cite{Andrews2010} suggests an outer radius of 10-15AU based on SMA observations at 880 $\rm \mu m$ and models the SED with an inner disc extending from 2 to 3.5 AU, and a second one extending from 3.5 to $\rm \sim 15~AU$.

Our best model for \object{TWA~04B} consists of a population of dust grains at 180 K placed at 4.8 AU and a second population of dust grains at 102 K placed at 7.9 AU from the star, with a total dust mass of  $\rm \sim 0.26~M_{\oplus}$. Our deduced radius for the inner disc is compatible within the errors with the value proposed by \cite{Andrews2010} while the radius of the outer disc is larger in our model. Our dust mass is two times higher than the value by \cite{Nilsson2010}, but we note that they used a single blackbody at 155 K to model the whole SED. Also, their flux density at 870 $\mu \rm{m}$ is lower than flux densities at similar wavelengths  from the literature.

\begin{figure}[!ht]
\begin{center}
   \centering
     \includegraphics[scale=0.35,angle=90]{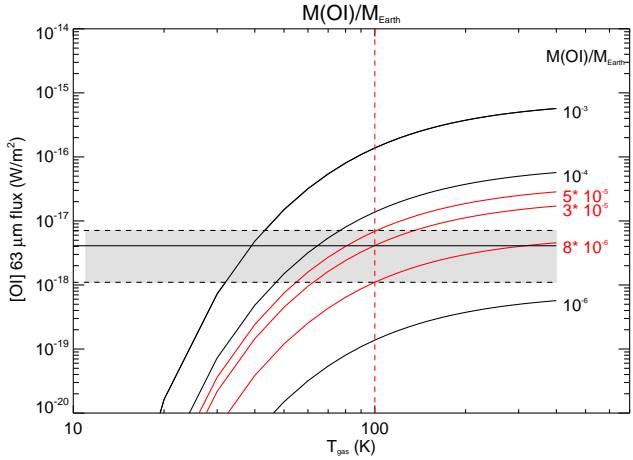} 
     \caption{Line luminosity versus gas excitation temperature for TWA 04B. The black and red curves represent the behaviour for different gas masses. The red ones show masses compatible with the observed flux range in TWA 04B for $\rm T_{gas}=100~K$. The grey dashed region shows the flux density values consistent with our observations.}
   \label{TWA04_OImass}
\end{center}
\end{figure}

Since we detected the [OI] emission in this system, we can study the gas content of the disc. In Fig. \ref{TWA04_OImass}, we show the line luminosity versus the excitation temperature for different gas masses. Assuming an excitation temperature of $\rm \sim$100 K for the [OI] line at 63.18 $\mu \rm{m}$, we can constrain the atomic oxygen gas mass to be between $\rm 8.0 \times 10^{-6} M_{\oplus}$ and $\rm 5.0 \times 10^{-5}~M_{\oplus}$. Assuming a primordial gas abundance for oxygen, i.e. $\rm 8.5 \times 10^{-4}$ with respect to H nuclei, the total gas mass ranges from $\rm 9.5 \times 10^{-3}~M_{\oplus}$ to $\rm 6.0 \times 10^{-2}~M_{\oplus}$,  implying a gas-to-dust ratio that ranges roughly from $\rm 3.6 \times 10^{-2}$ to 0.23. We note that, depending on the origin of the gas, the oxygen abundance can be larger, resulting in even lower gas masses and gas-to-dust ratios. Thus, we are detecting gas in a very low ratio to dust, so the system seems to be in a very advanced evolutionary stage. We can compare the OI gas detection in TWA 04B with those in TWA 01 \citep{Thi2010} and \object{HD172555} \citep{Riviere2012B}, because they share similar evolutionary stages (the general agreement is that the Beta Pictoris Moving Group is slightly older than TWA, with an absolute age in the range 12 to 20 Myr), although we must caution that \object{HD~172555} has a very different spectral type (A0 versus M). The result of this comparison is shown in Table \ref{GasComp}. TWA 04B shows the lowest ratio with respect to the 70 $\mu \rm{m}$ continuum emission. The gas in TWA 01 is probably primordial, while some authors \citep{Lisse2009,Johnson2012} argue that the gas in HD 172555 comes from the sublimation of silicates (olivines, pyroxenes) in hypervelocity collision. Considering the low line-to-continuum flux ratio in TWA 04B when compared with TWA 01, we suggest that the gas should have a second-generation origin, possibly from cometary evaporation. 

\begin{table}
\caption{Gas detections in TWA and Beta Pictoris moving group (BPMG)}             
\label{GasComp}      
\centering          
\begin{tabular}{lcll}     
\hline\hline       
Target & $\rm F_{[OI]} \over \lambda_{70 \mu m} \times F(\lambda_{70 \mu m})$& $\rm M_{dust}$ & $\rm M_{gas}$ \\
\hline
--  & -- & $\rm M_{\oplus}$ & $\rm M_{\oplus}$ \\
\hline   
TWA 04B & $\rm 1.5 \times 10^{-5}$ & 0.26 & $\rm (0.95-6) \times 10^{-2}$ \\
TWA 01& $\rm 2.2 \times 10^{-4}$ & 63 & 160-1600  \\
HD 172555 & $\rm 5.5 \times 10^{-3}$ & $\rm 4.8 \times 10^{-4}$ & $\rm 4.8 \times 10^{-4}$$^{(1)}$  \\
\hline
TWA 03A& $\rm < 2.0 \times 10^{-4} $ & 0.38 & $\rm < 0.042$$^{(2)}$ \\
TWA 07 & $\rm < 1.3 \times 10^{-3} $ & 0.15 & $\rm < 0.027$$^{(2)}$\\
TWA 11A & $\rm < 2.1 \times 10^{-5} $ & 0.15 & $\rm < 0.082$$^{(2)}$ \\
\hline                  
\end{tabular}
\tablefoot{(1): atomic oxygen mass only, see \cite{Riviere2012B}. (2): computed assuming primordial gas abundance for oxygen, i.e. $\rm 8.5 \times 10^{-4}$ with respect to H-nuclei.}
\end{table}

\cite{Soderblom1996} proposed a chromospheric origin for the $\rm{H}_{\alpha}$ emission, as we did in section \ref{Acc}. \cite{Dent2005} did not detect CO emission in \object{TWA~04B}. That \object{TWA~04B} shows [OI] in emission at 63 $\mu \rm{m}$ while showing no accretion would be pointing to a disc with an inner gap empty of gas, but with measurable amounts of gas at a larger radius. The origin of the gap may be related to the binary nature of the system, with separation of $\rm \sim 1.0 AU$.

\subsection{\object{TWA~07}} \label{TWA07}
\object{TWA~07} is an M1 star \citep{Webb1999} placed at 38 pc from the Sun. \cite{Low2005} used MIPS observations to study dust emission in \object{TWA~07} and modelled the SED with an 80 K blackbody at 6.8 AU from the star and with a minimum mass of $\rm \sim 4 \times 10^{-5}~M_{\oplus}$. Later on, \cite{Matthews2007} observations at 450 and 850 $\mu \rm{m}$ demonstrate that dust in a range of temperatures is needed to fit the 24 $\rm \mu m$ and the sub-millimetre flux densities simultaneously.

Fitting \object{TWA~07} with a single blackbody model results in a poor fit, while including a second blackbody highly increases the quality of the fit. The best model  consists of an inner disc at $\rm 66$ K, with $\beta \rm{=0.86}$ placed at $\rm \sim$ 38 AU from the star, and a second, very cold blackbody (20 K) at $\rm \sim$ 75 AU from the star. This is the coldest disc among the sample, and it is more than two times cooler than the Sun's Edgeworth-Kuiper Belt, at $\rm \sim 45~K$. As pointed out by \cite{Matthews2007}, no single temperature can fit both the near and mid-IR and the sub-millimetre flux densities. Including the 160 $\mu \rm{m}$ PACS photometric point pushes the outer blackbody to cooler temperatures and clearly demonstrates the bimodal shape of the SED. We are aware that the combination of many different grain size distributions can result also in a bimodal shape \citep[see][]{Matthews2007}, but we consider the presence of dust at different radii as the most plausible explanation. Because the flux density at 160 $\mu \rm{m}$ will underestimate the dust mass, we deduced a value of $\rm 0.15~M_{\oplus}$ using the flux density at $\rm{870} ~\mu \rm{m}$ by \cite{Matthews2007}.
 
According to section \ref{Acc}, the $\rm{H}_{\alpha}$ equivalent width agree with no ongoing accretion in \object{TWA~07}. The PACS line spectrum of this source does not show line emission. Both facts agree with an evolved, second-generation debris disc.

\subsection{\object{TWA~11A}}\label{TWA11A}
Infrared excess around \object{TWA~11A} was reported for the first time by \cite{Jura1991}, who highlighted the prominence of this excess when compared with $\beta$ Pictoris. \cite{Jura1993} studied the \object{TWA~11A} SED, and proposed a typical size of 10 $\mu \rm{m}$ for dust grains located at 40 AU from the star. \cite{Koerner1998} showed that the morphology of the emission at 20.8 $\mu \rm{m}$ agrees with a ring-like structure extending from 55 to 85 AU, while the excess at 12.5 $\mu \rm{m}$ arises predominantly from the region inside this radius, from gravitationally confined grains. Resolved images of  the disc have been studied by several authors \citep{Koerner1998,Jay1998,Schneider1999,Telesco2000}, who agree that the detected emission should arise from material at least at 55 AU. \cite{Augereau1999} modelled the SED with an outer disc at $\rm \sim$70 AU and an inner disc at $\rm \sim$9 AU, with a total solid mass of $\rm \sim 3.9 M_{\oplus}$ in a grain size distribution with $\rm a_{max}=100~cm$. Since the dust mass is $\rm M_{dust} \propto a_{max}^{0.5}$ in this model, the dust mass in grains $\rm < 1~mm$ is $\rm \sim 0.11~M_{\oplus}$. 

For TWA 11A,  including a second blackbody in the model does not result in a big improvement, so we kept a single blackbody model as the best option. The dust temperature is 108 K, corresponding to a disc at 59 AU, in good agreement with previous results with $\beta \rm{= 0.3}$. If we use the 160 $\mu \rm{m}$ flux density to derive the dust mass we get $\rm M_{dust}=0.27~ M_{\oplus}$, only two times the mass deduced by \cite{Augereau1999}, a difference that can be considered satisfactory given the uncertainties associated with the dust opacity law. We note that the model over-estimates the observed intensity at sub-millimetre wavelengths. A dust mass of $\rm \sim 0.146 ~M_{\oplus}$ was then computed using the flux density at 870 $\mu \rm{m}$ by \cite{Nilsson2010}, which agrees even better with previous results.

\object{TWA~11A} does not show  [OI] emission at 63 $\mu \rm{m}$. The lack of atomic emission in \object{TWA~11A} has already been reported and discussed in the study of atomic and molecular emission in the GASPS sample of HAeBe stars by \cite{Meeus2012}. The authors show how other stars with similar spectral types  emit in [OI] at 64.18 $\mu \rm{m}$ (e. g. \object{AB~Aur}, \object{HD~97048}). Therefore the lack of [OI] emission towards TWA 11A cannot be attributed to its spectral type. But we caution that discs around \object{AB~Aur} and  \object{HD~97048} are protoplanetary, gas-rich systems. Therefore, the lack of [OI] is not linked to the spectral type, but to the evolutionary stage of the system.

\section{Discussion}
TWA members detected with PACS show a wide variety of dust and gas properties from \object{TWA~01}, harbouring a protoplanetary-like disc with strong gas emission and ongoing active gas accretion, to \object{TWA~07}, which seems to be  devoid of gas but shows two debris rings at different radii. \object{TWA~04B} is considered a transitional disc, and it shows no accretion according to its $\rm{ H}_{\alpha}$ emission, but shows [OI] in emission at 63.18 $\mu \rm{m}$. Inner radii span a wide range of values, from $\rm \sim$ 0.4 AU for \object{TWA~03A} to 59 AU for \object{TWA~11A}, independent of the spectral type. The wide variety of properties in the five TWA discs studied most likely indicates that there are several factors driving disc evolution, and not just the age, including the dust/gas initial mass, the mass of the central object, angular momentum, multiplicity, composition of the original cloud, and possibly planet formation itself. Other authors have arrived at the same conclusion \citep{Furlan2006,Bayo2012,Lada2006,Hernandez2007,Currie2011}. 

\subsection{Dust in TWA}
TWA 01 shows the highest dust mass in the sample, $\rm \sim 63~M_{\oplus}$ \citep[][including PACS observations from GASPS]{Thi2010}, while other TWA members studied in the present work show dust masses in the range 0.15 to 0.38 $\rm M_{\oplus}$. According to our numbers, except for TWA 01, there is no mass left to form planets within the TWA members studied. The same result for a partially overlapping TWA sample was already reported by \cite{Weinberger2004}. Furthermore, we find no correlation between the spectral type and the dust mass; TWA 11A and TWA 07, which are A0 and M1 stars respectively, have approximately the same dust mass. On the other hand, TWA 01, which is an M2.5 star, shows a mass that is $\rm \sim 400$ times larger than TWA 07 mass. 

Another interesting point is the fact that TWA 11A, which is an A0 star \citep{Barrado2006}, does not show gas emission, while showing a very prominent IR excess. The disc is clearly a second generation debris disc, while TWA 01, TWA 03A and TWA 04B show SEDs that resemble those of transitional discs. This result agrees with the hypothesis that discs around early type stars tend to evolve faster toward the debris phase than late type stars. We note that we only have one early spectral-type object, and therefore this conclusion has to be taken with some caveats.

The $\beta$ values in the outer discs are typically around 0 (pure blackbody emission), with an average value of $\beta_{2} \rm{=0.12}$, which implies grains with sizes of the order of $\rm \sim$ 1000 $\mu \rm{m}$ \citep{Draine2006}, so significantly larger than ISM grains. This may be indicative of a second-generation origin for dust grains, but also may be a signpost of grain growth. On the other hand, $\beta$ values for inner discs have an average value of $\beta_{1} \rm{=0.53}$, i. e., similar to those values found  for T Tauri stars in Taurus-Auriga \cite[][with a typical value of 0.6]{Mannings1994}. This illustrates the need for populations with different sizes at different radii. 

The PACS detections in TWA can help us understand some evolutionary trends in the transition from a protoplanetary disc to a debris disc. The first is that all the detected objects in TWA show large IR excess, as already reported by \cite{Low2005} using \textit{Spitzer} data. There are no detections of weak excesses in the sample. This may suggest that the dust (and gas) depletion rate is faster than the replenishment rate by collisions between planetesimals from an unseen population. If discs are formed with an initially low gas and dust content, it may be that large planetesimals never formed. On the other hand, the high-mass objects detected in this survey may be those with enough initial gas and dust that a significant population of planetesimals exist.

\cite{Donaldson2012} have recently modelled the SED of some Tuc Hor members that were also part of the GASPS sample and concluded that all the detected discs were debris discs, with no [OI] emission. In the present study, we have detected five discs in TWA, but they span a wide range of disc parameters, including both debris discs and protoplanetary discs. Tuc Hor is a 30 Myr old association, while TWA is significantly younger (8 to 20 Myr, depending on different authors). The different range in ages can explain not only the presence of protoplanetary/transitional systems in TWA, but also the fact that we detect [OI] emission at 63 $\mu \rm{m}$ in two systems (namely TWA 01 and TWA 04 B), while no [OI] emission is observed in Tuc Hor.

\subsection{Gas in TWA}\label{TWAgas_sec}
There are some interesting questions regarding the gas content in TWA member discs. The [OI] detection in \object{TWA~01} (TW Hya) was studied in detail in \cite{Thi2010}, where they estimated a gas mass of 160 to 1600 $\rm M_{\oplus}$. This mass is four to six orders of magnitude higher than the mass that we estimated for \object{TWA04~B}. If we estimate the gas mass for \object{TWA~01} in the same way as we did for \object{TWA~04B}, we get a mass in the range  0.3 to 0.6 $\rm M_{\oplus}$, so much lower than the more realistic value by \cite{Thi2010}, but again 5 to 61 times higher than the gas mass in TWA 04B. Both stars have similar spectral-types: M2.5 for \object{TWA~01} \citep{Vacca2011} and M5 for \object{TWA~04B} \citep{Gregorio1992}. Given that the spectral types are so similar, what is driving the large difference in gas mass? One possibility is that the difference comes from the size of the disc, therefore the total initial mass available. The disc in \object{TWA~04B} is truncated by the primary, \object{TWA~04A}, at $\rm \sim 15 AU$, while \object{TWA~01} is expected to be much more extended \citep[196 AU,][]{Qi2004}. Also, \object{TWA~04B} is itself a close spectroscopic binary, which opens a large inner gap devoid of gas and dust; therefore, the difference in dust and gas evolution can also be related to multiplicity. 

\begin{figure}[!t]
\begin{center}
   \centering
     \includegraphics[trim = 31mm 0mm 0mm -20mm,clip,scale=0.35,angle=90]{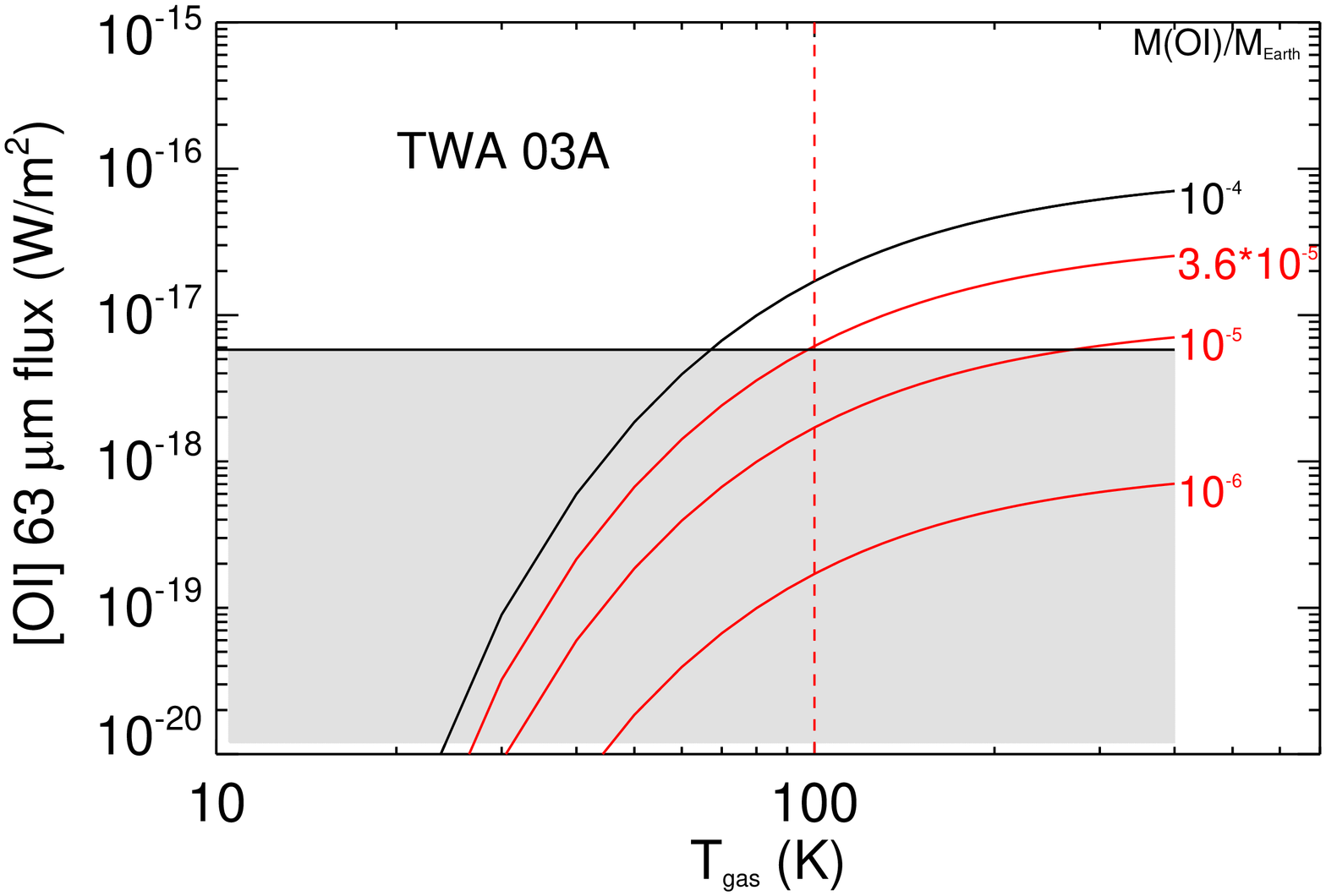} 
     \includegraphics[trim = 31mm 0mm 0mm -20mm,clip,scale=0.35,angle=90]{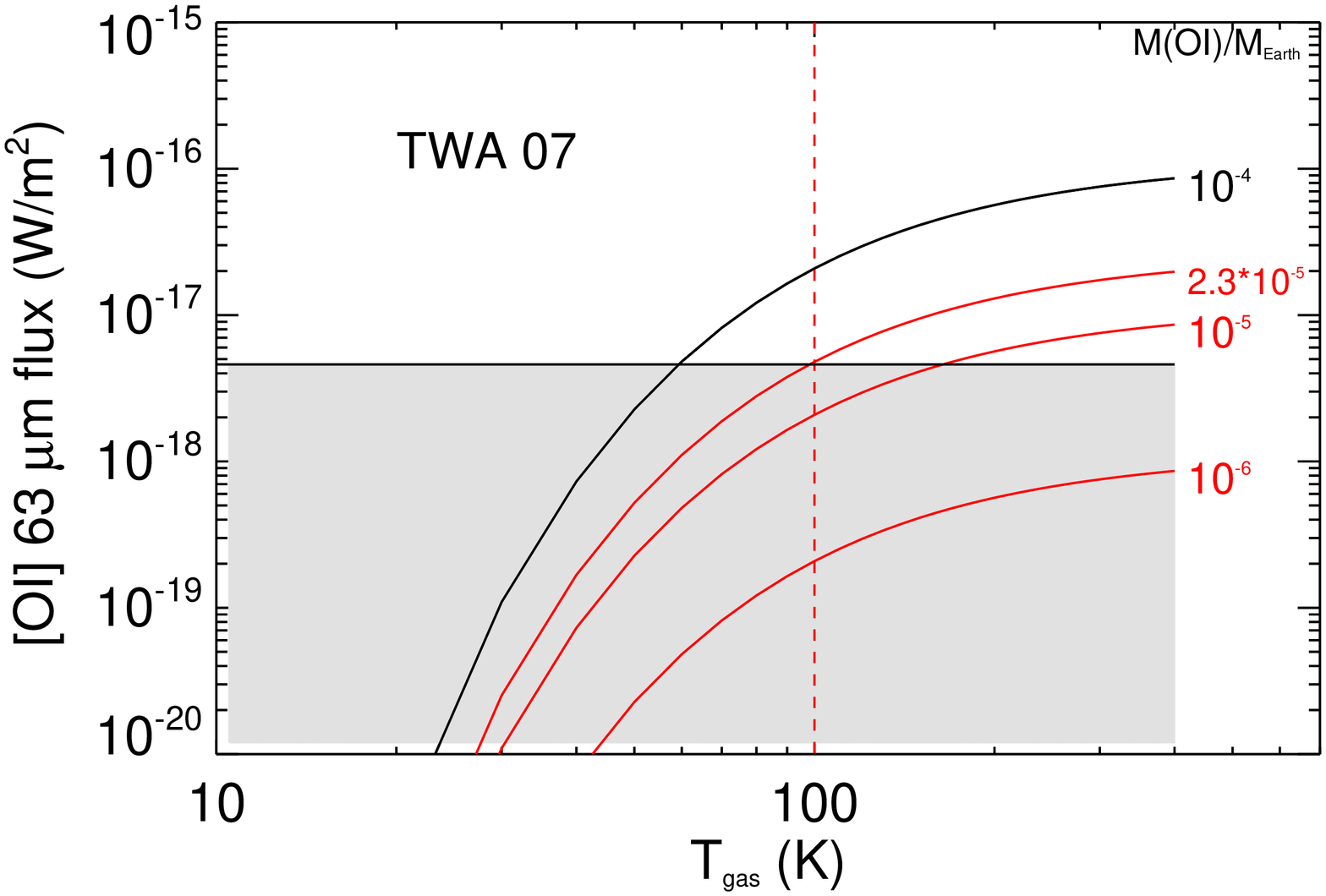} 
     \includegraphics[trim = 0mm 0mm 0mm -20mm,clip,scale=0.35,angle=90]{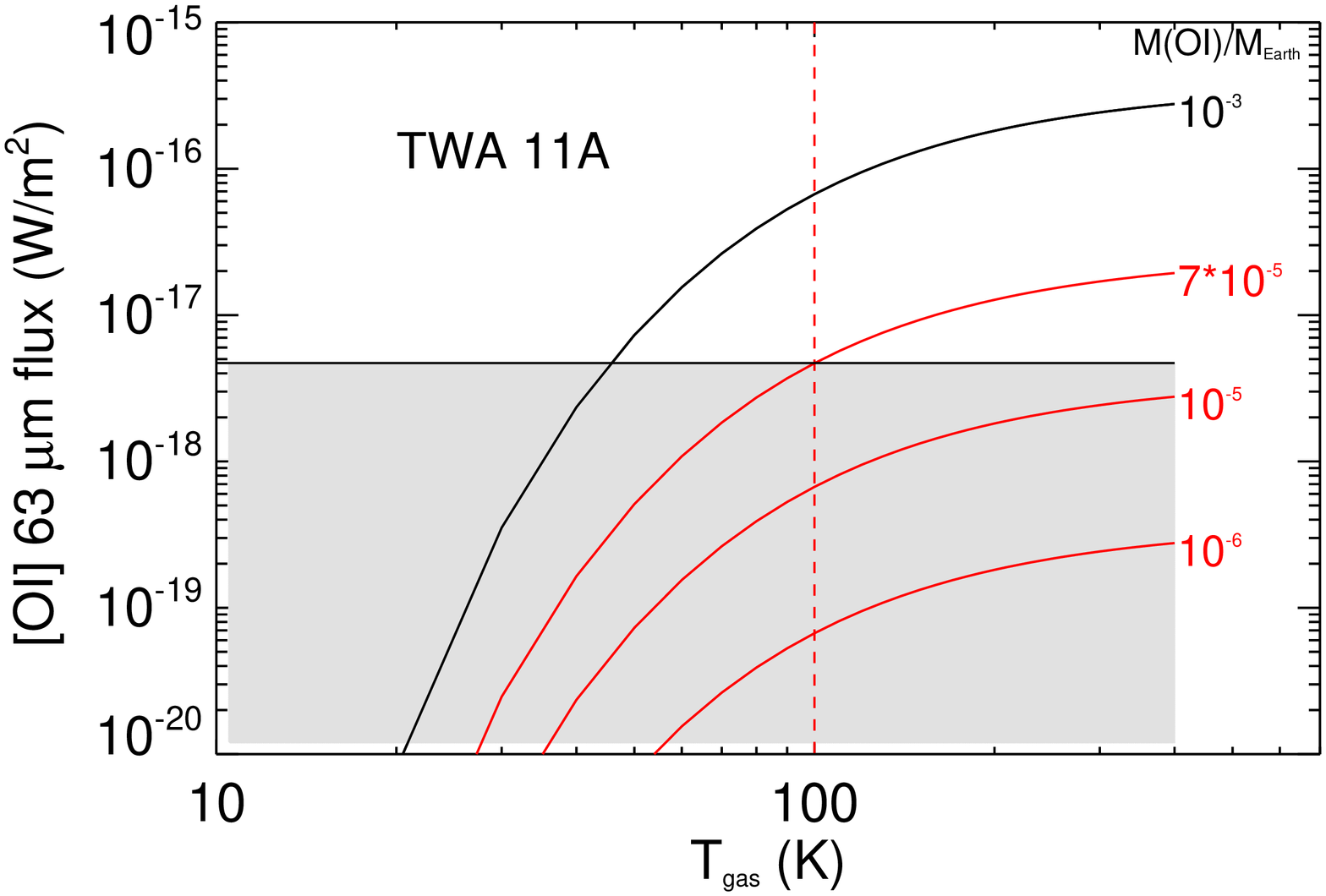} 
     \caption{Line luminosity versus gas excitation temperature for TWA members with no [OI] emission detected. The black and red curves represent the behaviour for different gas masses. The red ones show masses compatible with the flux upper limits for $\rm T_{gas}=100~K$. The grey dashed region shows the parameter space compatible with observations.}
   \label{TWAgas}
\end{center}
\end{figure}

In Fig. \ref{TWAgas} we show [OI] luminosity at 63 $\mu \rm{m}$ versus the excitation temperature for objects not detected with PACS at 63.18 $\mu \rm{m}$. As can be seen, upper limits on the [OI] gas mass are a few $\rm 10^{-5}~M_{\oplus}$, which is similar to the [OI] gas mass in \object{TWA~04B}. The gas mass in \object{TWA~01} is therefore more than one order of magnitude higher than any other gas mass in the association.

It is also interesting to compare the detection of [OI] gas in \object{TWA~04B} with the non-detection in \object{TWA~03}. In Sec. \ref{Acc} we classified \object{TWA~03} as an accreting disc and \object{TWA~04B} as a non-accreting disc. The [OI] gas detection combined with the absence of accretion in \object{TWA~04B} can be explained by the fact that \object{TWA~04B} is a close spectroscopic binary, and therefore no gaseous material could survive inside 1 AU, and no accretion could be detected \citep[see][]{Prato2001}. Our own determination of the inner radius in \object{TWA~04B} agrees with an inner hole $\rm >1~AU$.  

\begin{figure}[!t]
\begin{center}
   \centering
     \includegraphics[scale=0.50]{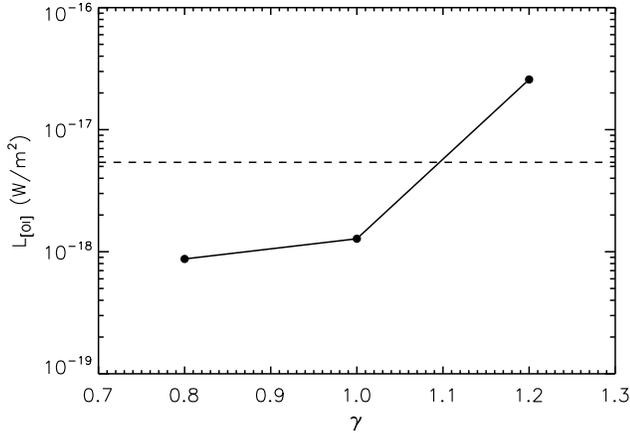} 
     \caption{[OI] luminosity at 63.18 $\mu \rm{m}$ versus flaring index ($\gamma$) for a model star with $\rm L_{*}=0.13~L_{\odot}$, $\rm R_{in}=R_{sub}$, $\rm R_{out}=100~AU$, $\rm a_{min}=0.05~\mu m$, $\rm a_{min}=1000~\mu m$, $\rm \epsilon = -1.0$, $\rm f_{UV}=0.1$ and $\rm M_{dust}=M_{gas}=10^{-6}~M_{\odot}$ at the distance of \object{TWA~03A}. The model is for illustration, and is not intended to represent TWA 03A exactly, but a generic system with similar stellar parameters. The horizontal, dashed line represents the average upper limit for objects in TWA.}
   \label{LOIvsBeta}
\end{center}
\end{figure}

That that we do not detect any [OI] emission in \object{TWA~03A} while we see signs of accretion is intriguing. A possible explanation relies on the geometry of the disc. If the disc is very flat, then there is not enough material irradiated by the central star, small amounts of oxygen are excited, and the resulting [OI] emission should be too weak to be detected. We do not claim that this is the only possible explanation. A lower ultraviolet (UV) flux could also explain the difference between \object{TWA~03A} and \object{TWA~04B}, since the UV flux is the main source of energy for the gas heating via photoelectric effects on dust grains and polycyclic aromatic hydrocarbons. However, both targets are a similar spectral type so we expect that the UV flux is very similar. Indeed, \object{TWA~03A} is actively accreting, and therefore we expect a higher UV flux from that source, so a difference in the flaring geometry is the most plausible explanation for the difference in [OI] emission. In Fig. \ref{LOIvsBeta} we represent the [OI] luminosity at 63.18 $\mu \rm{m}$ versus the flaring index ($\gamma$) for a model system taken from the DENT grid of models \citep{Kamp2011} scaled to the distance of \object{TWA~03A}, 42 pc.  As can be seen, the flux for the same star dramatically changes by more than one order of magnitude by changing the flaring index, in such a way that only the system with $\gamma \rm{=1.2}$ could be detected by our PACS observations, while all of them share the same gas and dust mass. Overall, Fig. \ref{LOIvsBeta} shows that in discs with low-mass gas, hence a low mass-accretion rate, OI is very difficult to detect with current instrumentation. 

In Taurus, 46 stars with discs out of 76 observed show [OI] emission \citep[][submitted]{Howard2013}, while in TWA we detected line emission in two systems out of the nine observed and in two out of the five known to have discs. Although the number of observations or detections in TWA is too low to make statistical arguments, we highlight the difference in gas-rich discs fractions from Taurus (0.6) to TWA (0.4). \cite{Riviere2012} showed that warm water emission is commonly found among Taurus T Tauri stars, but we do not detect warm water emission toward TWA members; while $\rm \sim 24 \% $ of the gas-rich Taurus members (i. e., those showing [OI] emission at 63 $\mu \rm{m}$) show water emission at 63.32 $\mu \rm{m}$, no TWA members show water emission at 63 $\mu \rm{m}$. According to \cite{Riviere2012}, the water emission around T Tauri stars in Taurus comes from a region at $\rm \sim 1~AU$ and $\rm \sim 3~AU$ wide. TWA 04B, TWA 07, and TWA 11A show inner radii larger than 3 AU, therefore no water could be detected in the regions. For TWA 01 and TWA 03A, the lack of warm water emission must have a different explanation, most probably one related to reprocessing of circumstellar gas or the geometry of the discs.

\section{Summary and conclusions.}
We observed 14 TWA members with the PACS instrument on board the \textit{Herschel Space Observatory}. All of them were observed with the PACS photometer (70, 100, and/or 160 $\mu \rm{m}$). Nine were observed with PACS Line Spectroscopy, targeting [OI] at 63 $\mu \rm{m}$. The main conclusions are the following:

1. We detected excess photometric emission at 70 $\rm \mu m$, 100 $\rm \mu m$ and 160 $\mu \rm{m}$ in 4,4 and 5 systems out of 12, 9 and 14 observed at these wavelengths. We detected for the first time 100 $\rm \mu m$ and 160 $\mu \rm{m}$ emission towards TWA 07, and also 100 $\mu \rm{m}$ emission towards TWA 03A. Objects not detected at 70 $\mu \rm{m}$ show upper limits near the photospheric level. Therefore, if present, any circumstellar material must be cold or very low mass.
 
2. Among the five systems detected with IR excess, two of them (TWA 01 and TWA 04B) show [OI] emission at 63.18 $\mu \rm{m}$, indicating the presence of gas in those systems. None of the systems show water emission at 63.32 $\mu \rm{m}$. Future research is needed to understand the gas emission in TWA 04B.

3. We modelled the dust IR emission with blackbody models, and used them to derive dust masses and inner radii, thereby providing  temperatures in the range 20 (TWA 07) to 280 K (TWA 04B). Dust masses are in the range 0.146 (TWA 07 and TWA 11A, from blackbody models) to 63 $\rm M_{\oplus}$ (TWA 01, from detailed modelling). Disc radii are in the range 0.4 to 59 AU.

TWA members show a wide variety of disc properties, implying different stages of disc evolution, from the protoplanetary/transitional, gas-rich disc around TWA 01 to the very cold, gas-free debris disc surrounding TWA 07. We propose that there must be several factors (others than the age) driving the evolution of the gas and dust contents in circumstellar environments, such as multiplicity, disc mass, stellar mass, angular momentum and composition, and, in general, initial conditions.

\begin{acknowledgements}
This research has been funded by Spanish grants AYA
2010-21161-C02-02, AYA2012-38897-C02-01, AYA2011-26202 CDS2006-00070, and PRICIT-S2009/ESP-1496. We also
acknowledge support from ANR (contract ANR-07-BLAN-0221) and
PNPS of CNRS/INSU, France. C. Pinte acknowledges funding from the 
European Commission's 7$^\mathrm{th}$ Framework programme 
(contract PERG06-GA-2009-256513) and from 
Agence Nationale pour la Recherche (ANR) of France under contract
ANR-2010-JCJC-0504-01. WFT thanks the CNES for a post-doctoral position. WFT, FM and IK acknowledge
funding from the EU FP7-2011 under Grant Agreement nr. 284405 (DIANA
project). JCA acknowledges the PNP-CNES for financial support. FM acknowledge support from the Millennium Science Initiative (Chilean Ministry of Economy), through grant ``Nucleus P10-022-F''. 
\end{acknowledgements}

\bibliographystyle{aa} 
\bibliography{biblio.bib}
\end{document}